\documentclass{article}
\usepackage{listings}
\usepackage{color}
\usepackage{tocloft}
\usepackage[margin=1.1in]{geometry}
\usepackage{times}
\usepackage{bm}
\usepackage{natbib}
\newtheorem{thm}{Theorem}[section]
\newtheorem{example}[thm]{Example}
\usepackage{afterpage,lipsum,capt-of,graphicx}
\usepackage{graphicx}
\usepackage{subcaption}
\usepackage{amsfonts,amsmath}
\usepackage{url}
\usepackage{booktabs} 
\usepackage{prodint}
\usepackage{bbm}
\usepackage{bm}
\usepackage{tabularx}
\usepackage{enumitem}
\newcolumntype{L}{>{\raggedright\arraybackslash}X}
\usepackage{color}
\newcommand{\R}{\mathbb{R}}
\newcommand{\N}{\mathbb{N}}
\newcommand{\1}{\mathbbm{1}} 

\newcommand{\EE}{\mathbb{E}}
\newcommand{\eps}{\varepsilon}
\newcommand{\cupdot}{\mathbin{\mathaccent\cdot\cup}}
\newcommand\independent{\protect\mathpalette{\protect\independenT}{\perp}}\def\independenT#1#2{\mathrel{\rlap{$#1#2$}\mkern2mu{#1#2}}}

\usepackage[dvipsnames]{xcolor}
\definecolor{afb}{rgb}{0.36, 0.54, 0.66}

\def\bSig\mathbf{\Sigma}

\title{Ranking of average
  treatment effects with generalized random forests for
  time-to-event outcomes}

\author{\small Helene C. W. Rytgaard$^{1,*}$, 
Claus T. Ekstr{\o}m$^{1}$, Lars V. Kessing$^{2}$, and Thomas A. Gerds$^{1}$ \\
\small $^{1}$Section of Biostatistics,
  University of Copenhagen,
  Oester Farimagsgade 5, Copenhagen
Denmark\\
\small$^{2}$Copenhagen Affective Disorder research Center
(CADIC), \\ \small Psychiatric Center Copenhagen, Rigshospitalet,
University of Copenhagen}

\begin{document}

\date{\it January 28, 2021} 

\label{firstpage}

\begin{abstract}
  In this paper we present a data-adaptive estimation procedure for
  estimation of average treatment effects in a time-to-event setting
  based on generalized random forests.  In these kinds of settings,
  the definition of causal effect parameters are complicated by
  competing risks; here we distinguish between treatment effects on
  the crude and the net probabilities, respectively.  To handle
  right-censoring, and to switch between crude and net probabilities,
  we propose a two-step procedure for estimation, applying inverse
  probability weighting to construct time-point specific weighted
  outcomes as input for the forest. The forest adaptively handles
  confounding of the treatment assigned by applying a splitting rule
  that targets a causal parameter.  We demonstrate that our method is
  effective for a causal search through a list of treatments to be
  ranked according to the magnitude of their effect. We further apply
  our method to a dataset from the Danish health registries where it
  is of interest to discover drugs with an unexpected protective
  effect against relapse of severe
  depression. \\
\end{abstract}

\maketitle

\section{Introduction}
\label{sec:introduction}

Drug repurposing is an important low-cost method for drug discovery
which is typically based on a data-driven experimental approach. In
this paper, our general aim is the ability to rank a list of treatment
variables according to their effect on a time-to-event outcome.  We
consider average treatment effect estimation based on generalized
random forests in a time-to-event setting with competing risks.  We
find two aspects particularly important when having to search through
a potentially large list of treatments. First, the search algorithm
should be as flexible as possible: In drug repurposing studies, in
particular, one may have expert knowledge of the outcome process being
studied but not the treatments, and if one has many treatments to
search through it will be impossible to correctly specify parametric
models for all treatment propensities with main effects and
interactions. Second, we need a real-valued measure to be used for
ranking that should have a sensible interpretation.  Compared to other
methods for average treatment effect estimation in right-censored and
competing risks settings \citep[see, e.g.,][]{ozenne2020estimation},
the methods presented in this paper do not require specification of
models for treatment propensity and outcome distribution. Further, we
discuss the choice between different causal parameters in the
competing risks setting when the aim is to identify new active
substances.

A random forest \citep{breiman2001random} is a popular data-driven
algorithm that can be used for variable importance analysis, i.e., to
rank variables according to their association with the outcome of
interest \citep{ishwaran2007variable,strobl2008conditional}. Mostly,
these variable importance measures are based on prediction performance
\citep{breiman2001random,ishwaran2007variable} or based on the tree
building process of the forests
\citep{ishwaran2010high,ishwaran2011random}. Our approach in this
paper is different in that we consider the use of causal treatment
effect parameters as a variable importance measure. Similar approaches
have also been considered in the context of high-dimensional biomarker
discovery, see, for example,
\citet{tuglus2008targeted,bembom2009biomarker,wang2011dimension}.  In
a counterfactual framework
\citep{neyman1923applications,rubin1974estimating}, treatment effect
parameters are formally defined as a difference between expected
counterfactual outcomes. Under a set of structural and distributional
assumptions the parameters are linked to the observed data. We
formulate causal parameters in terms of average differences of event
probabilities at pre-specified time horizons of interest, allowing us
to report a time-point specific measure of the effect of a particular
treatment.

Generalized random forests (GRFs)
\citep{wager2018estimation,athey2019generalized} are a recent
extension of Breiman's random forests that have been applied to
provide nonparametric inference for heterogeneous treatment effects in
settings with real-valued and uncensored outcomes of interest. The GRF
algorithm is implemented to optimize estimation of the causal
treatment effect specifically.  Here we implement GRFs for
time-to-event outcomes by using inverse probability weighting to make
the GRF implementation directly applicable to our setting with
right-censoring and competing risks.  In the competing risks setting,
we further discuss the distinction between treatment effects on crude
and net probabilities. These considerations are closely related to the
work of \citet{young2020causal}.

Our motivation comes specifically from a large-scale observational
registry study on drug purchases and development of psychiatic
disorders. Here the goal is to discover if drugs that are already in
clinical use may have a protective effect against depression.
Psychiatric disorders is a field where the pharmaceutical industry has
substantially withdrawn from developing new drugs; thus, in the
absence of new randomized clinical trials, and to supplement the
expensive and time-consuming generation of data from clinical trials,
a systematic search through all drug purchases in the registry data is
a cost-efficient way to identify new treatments as well as to discover
adverse side-effects.  Specific findings can then subsequently be
further investigated in randomized trials. For proof of concept and
illustration, we analyze Danish registry data on all Danish citizens
who have a first time diagnosis with depression registered. We follow
these patients until depression relapse, onset of other mental
disorders, death without relapse, or right-censoring, and apply our
proposed method to rank drug treatments according to the magnitude of
their effect on depression relapse.

The article is organized as follows.  In Section
\ref{sec:setting:notation} we introduce the setting and notation for
survival and competing risks data. We define our target parameters in
terms of counterfactual outcomes, and we discuss the distributional
assumptions under which we can identify the parameters from the
observed data. In Section \ref{sec:grf:with:pseudo} we review the
generalized random forest methodology and present our weighting
approach for making the methodology applicable to time-to-event data.
In Section \ref{sec:variable:importance} we introduce and discuss the
use of average treatment effects specifically for the purpose of
variable importance analysis\color{black}. In Section
\ref{sec:simulation:strategy} we study the performance using simulated
data. In Section \ref{sec:data:application} we analyze Danish registry
data.  We close with a discussion in Section \ref{sec:discussion}.

\section{Setting and notation}
\label{sec:setting:notation}

In time-to-event settings subjects are observed from study entry to
the occurrence of an event of interest or a competing event. If no
event of any kind is observed within the subject-specific follow-up
time, the subject is right-censored. Specifically, we consider a
competing risks situation with \(J\ge 2\) mutually exclusive types of
events. For sake of presentation, we assume throughout that
\(J=2\). We denote by \(T_i\) the uncensored event time, by
\(\Delta_i\in\lbrace 1, 2\rbrace\) the event type and by \(C_i\) the
censoring time, such that the observed data are
\(\tilde{T}_i = \min (T_i, C_i)\) and
\(\tilde{\Delta}_i = \1 \lbrace T_i \le C_i\rbrace
\Delta_i\). Moreover, \(\bm{X}_i \in \mathcal{X} \subseteq \R^p\) is a
vector of baseline covariates and
\(\bm{A}_i = (A_{1,i}, \ldots, A_{K,i})\in \lbrace 0, 1\rbrace^K\) is
a vector of \(K\in\N\) binary treatment variables with \(A_k=1\)
indicating treatment and \(A_k=0\) no treatment, \(k=1, \ldots, K\).
The data consist of \(n\in \N\) independent samples,
\( \big\lbrace (\bm{X}_1, \bm{A}_{1},\tilde{T}_1,\tilde{\Delta}_1),
\ldots, (\bm{X}_n, \bm{A}_{n},\tilde{T}_n,\tilde{\Delta}_n)
\big\rbrace\). We are interested in estimating the effect of the
treatment variable \(A_k\in\lbrace 0,1\rbrace\) on the probability of
events of type \(j=1\). We refer to the other type of events
(\(j= 2\)) as competing events, or competing risks.  We define our
target parameter in terms of counterfactuals, using a notation with
superscripts to define interventions. In particular, we define \(T^a\)
as the uncensored counterfactual event time and \(\Delta^a\) as the
corresponding event indicator that would result from setting treatment
\(A_k\) to \(a\). Further, for \(j=1,2\), we use \(T^{j,a}\) to denote
the uncensored counterfactual event time of type \(j\) that would
result if treatment \(A_k\) had been set to \(a\) in a hypothetical
world where cause \(j\) is the only cause.  Note that we distinguish
between the counterfactual event time variable \(T^a\) with a single
superscript and the counterfactual event time variable \(T^{j,a}\)
with double superscript.  Note also that when studying the treatment
\(A_k\), the other treatments can enter the vector of baseline
covariates.

\subsection{Treatment effects in presence of competing risks}
\label{sec:target:parameter}

\subsubsection{The competing risks problem revisited}
\label{sec:competing:risks:problem}

As a motivation for our later discussions on causal parameters for
treatment effect ranking, we here briefly revisit the problems with
causal inference in competing risks settings. In particular, in
presence of competing risks, the one-to-one correspondence between the
cause-specific hazard and the absolute risk is lost
\citep{andersen2012competing}, and the effect of variables on the
cause-specific hazard may be quite different from their effect on the
absolute risk \citep{gray1988class}. Specifically, variables may have
an indirect effect on the absolute risk only through its effect on the
cause-specific hazard of the competing event. Consider the following
example.

\begin{example}
  Suppose that it is of interest to rank two treatments, \(A_1\) and
  \(A_2\), according to their effect on the event of interest.  Assume
  that the cause-specific hazard rates are given as follows for the
  event of interest (\(\lambda_1\)) and the competing event
  (\(\lambda_2\)):
  \begin{align*}
    \lambda_1 ( t \, \vert \, A_1, A_2) &= e^{-0.2A_1 - 0.2A_2}, \quad \text{and,} \quad
    \lambda_2 ( t \, \vert \, A_1, A_2) = e^{-0.2A_1 }. 
  \end{align*}
  Clearly, \(A_1\) and \(A_2\) have the same effect on the hazard of
  the event of interest. Nonetheless, the cause-specific cumulative
  incidence of the event of interest also depends on the hazard rate
  of the competing event.  Now, assume that \(A_1\) and \( A_2\) are
  both Bernoulli variables with \(P(A_k=1)=0.5\), \(k=1,2\). The
  average effects at time \(t=0.1\) are very similar,
  \( \EE [ F_1 (0.1 \mid 0.1, A_2 ) - F_1 (0.1 \mid 0, A_2)]=
  -0.0138\) and
  \(\EE [ F_1 (0.1 \mid A_1, 1 ) - F_1 (0.1 \mid A_1, 0)] = -0.0145\),
  whereas for increasing \(t\) they get very different: At time
  \(t=1\), for example, we have
  \( \EE [ F_1 (1\mid 1, A_2 ) - F_1 (1 \mid 0, A_2)] = -0.0290 \) and
  \( \EE [ F_1 (1 \mid A_1, 1 ) - F_1 (1 \mid A_1, 0)] = - 0.0547 \),
  i.e., a considerably larger effect of \(A_2\).  Thus, in this
  example, due solely to the effect that \(A_1\) has on the competing
  cause-specific hazard rate, we would conclude very different effects
  of the two treatments on the cumulative risk of cause 1.
  \label{ex:hazard:vs:risk}
\end{example}

Our goal is variable importance and the ability to rank a list of
treatment variables according to their effect on a specific
time-to-event outcome. Example \ref{ex:hazard:vs:risk} illustrates the
interpretational issues with absolute risks in the presence of
competing risks, and the question is if we would like to conclude
different effects for the two treatments \(A_1\) and \(A_2\). This
problem is not solved by analyzing the cause-specific hazard rates
alone. Specifically, these are defined conditional on post-treatment
mechanisms and therefore cannot be ascribed an interpretation as a
measure of a causal treatment effect
\citep{hernan2010hazards,martinussen2020subtleties}.

In the following, we discuss the distinction between effects on
\textit{crude} and \textit{net} probabilities, respectively, to
characterize the effect of a treatment variable \(A_k\) on the
occurrence of events of type \(j=1\). We emphasize that the choice
between crude and net effects corresponds to the choice between
different causal parameters, and altogether depends upon the goal of
the analysis. In summary, we argue that:
\begin{enumerate}[align=left,labelsep=0.4pt]
\item[1. ] Causal effects on \textbf{crude probabilities} are used for
  describing the real world; crude probabilities allow us to infer on
  treatment effects that would actually occur in a given
  population.\vspace{0.2cm}
\item[2. ] Causal effects on \textbf{net probabilities} are defined in
  hypothetical worlds without competing risks and reflect effects of
  etiological nature.  They allow us to infer treatment effects
  directly on the event type of interest without interference from
  indirect effects on the competing event time.
\end{enumerate}
Different assumptions on the underlying data-generating mechanisms are
necessary when focus is on crude or on net probabilities as we
describe in Section \ref{sec:identifiability:crude}.  Importantly, the
assumptions needed to identify net probabilities are considerably more
ambitious and net probabilities have thereby been criticized
\citep{andersen2012interpretability}. In this work we argue that in a
variable importance analysis (Section \ref{sec:variable:importance}),
the subject matter interest may not in the treatment effects on the
crude probability scale but rather to assess treatment effects only
directly on the occurrence of type \(j = 1\) events.  In Section
\ref{sec:crude:target:parameter}, we start by discussing treatment
effects on crude probabilities. In Section
\ref{sec:net:target:parameter}, we present treatment effects on net
probabilities.

\subsection{Effects on crude probabilities}
\label{sec:crude:target:parameter}

Recall that the random variables \(T^0\) and \(T^1\) denote the
uncensored counterfactual event times that would result if treatment
had been set to \(A_k=0\) or \(A_k=1\), respectively.  The average
treatment effect (ATE) of \(A_k\) on the crude risk of events of type
1 before a fixed time horizon \(t_0 >0\) is defined as follows
\begin{align}
\begin{split}  \bar{\theta}_{\mathrm{crude}} &=
                                  P( T^1 \le t_0, \Delta^1=1 ) -
                                  P(  T^0  \le t_0, \Delta^0=1 ).
                                \end{split}\label{eq:psi2:bar}
\end{align}
The quantities \(P( T^a \le t_0, \Delta^a=1 )\), \(a=0,1\), in
\eqref{eq:psi2:bar}, referred to as the crude probabilities, are the
cumulative incidence functions \citep{gray1988class} of the event of
interest for a hypothetical treated and a hypothetical untreated
population, respectively. These crude probabilities also depend on the
hazard rate of the competing event, since, at any time, the event of
interest can only occur for subjects who have survived all risks so
far. A treatment which reduces the hazard rate of the competing risk
increases the event-free survival probability and thereby indirectly
increases the crude risk of the event of interest, and vice
versa. Particularly, a treatment effect reflected in a non-zero value
of \(\bar{\theta}_{\mathrm{crude}}\) will occur also if there is only
an indirect effect via the hazard rate of the competing event.

\subsection{Effects on net probabilities}
\label{sec:net:target:parameter}

Recall that the counterfactual random variables \(T^{1,0}\) and
\(T^{1,1}\) are the (uncensored) counterfactual event times that would
have been observed in a hypothetical world in which cause \(j=1\) is
the only cause and where treatment had been set to \(A_k=0\) and
\(A_k=1\), respectively. Particularly, \(T^{1,0}\) and \(T^{1,1}\) are
latent times that are not always observed in the real world due to
cause \(j=2\) events and due to right-censoring.  The average
treatment effect of \(A_k\) on the net risk of events of type 1 is
defined as follows
\begin{align}
  \bar{\theta}_{\mathrm{net}} & =
                   P( T^{1,1} \le t_0  ) - P( T^{1,0} \le t_0  ).  \label{eq:psi1:bar} 
\end{align}
We emphasize that, opposed to the crude risks
\(P( T^a \le t_0, \Delta^a=1 )\), \(a=0,1\), in Equation
\eqref{eq:psi2:bar}, the net risks \(P( T^{1,a} \le t_0 )\),
\(a=0,1\), are not affected by the (indirect) effect that a treatment
may have on the hazard rate of the competing risk. They are
interpreted as net probabilities for the event of interest in a
hypothetical world where the competing event cannot happen. A
treatment effect reflected in a non-zero value
\(\bar{\theta}_{\mathrm{net}}\) will only occur if the studied
treatment has a direct effect on the event of interest.

\subsection{Identifiability of treatment effects on crude and net
  probabilities}
\label{sec:identifiability:crude}
\label{sec:identifiability:net}

The average treatment effects on crude probabilities
\(\bar{\theta}_{\mathrm{crude}} \) and net probabilities
\(\bar{\theta}_{\mathrm{net}} \) are defined in terms of
counterfactual random variables, and are identified from the observed
data only under causal assumptions \citep{hernanrobins}.  We review
these assumptions in the supplementary material (Appendix A)
separately for \(\bar{\theta}_{\mathrm{crude}} \) and
\(\bar{\theta}_{\mathrm{net}} \).  We here point out the assumption of
\textit{no unmeasured confounding} only, to really contrast the choice
between the two parameters. Particularly, for identifiability of the
crude effects, this is an assumption of conditional independence
between the counterfactuals and the treatment and censoring
mechanisms, as follows,
\( (T^a, \Delta^a) \independent A_k \, \vert \, \bm{X}\), for
\(a=0,1\), and
\( (T, \Delta) \independent C \, \vert \, A_k, \bm{X}\). To move from
crude to net effects, one needs additionally that
\(T^{1,A_k} \independent T^{2,A_k} \, \vert \, A_k, \bm{X}\).
As previously mentioned, we stress that this is a very strong
assumption: Whether \(A_k\) and \(\bm{X}\) together include all
factors that we believe to be predictive of both event types depends
very much on the nature of the competing events and how rich the
measured set of covariates is.

\section{Generalized random forests with inverse probability weighted
  outcomes}
\label{sec:grf:with:pseudo}

Generalized random forests (GRFs) \citep{athey2019generalized} are a
recent generalization of the original random forest algorithm
\citep{breiman2001random}, a machine learning tool that adaptively
searches the covariate space by recursive sample splitting. Generally,
a forest consists of \(B\in\N\) randomized trees, where the \(b\)th
tree of the forest is grown by recursively splitting the covariate
space according to some split criterion.  GRFs provide a data-adaptive
approach to estimation of conditional treatment effects for uncensored
data, particularly, for a generic outcome variable \(Y\in\R\),
\(\theta(\bm{x})= \EE [Y \, \vert \, A_k=1, \bm{X}=\bm{x}] - \EE [Y \,
\vert \, A_k=0, \bm{X}=\bm{x}]\).  A key part of the generalized
random forest algorithm is the splitting rule that targets
specifically the estimation of the quantity \(\theta(\bm{x})\) of
interest; particularly, each tree applies a splitting rule that
adaptively makes binary partitions of the covariate space such as to
maximize heterogeneity in \(\theta(\bm{x})\).  By averaging over
neighborhoods defined by the trees, the forest produces a neighborhood
function that is used as a kernel for estimation of
\(\theta(\bm{x})\). In the supplementary material (Appendix C) we
describe the local gradient-based criterion for making splits and the
kernel-based estimator for average treatment effects for uncensored
data as proposed by \cite{athey2019generalized}. We further review the
structural model formulation of treatment effects of \citet[][Section
6]{athey2019generalized} and its relation to our setting with the
counterfactual formulation.

The problem in our setting is that we do not observe the actual
outcomes of interest. For the parameter
\(\bar{\theta}_{\mathrm{crude}}\), for example, we do not observe
\(Y:= \1 \lbrace T \le t_0, \Delta =1 \rbrace\) due to
right-censoring. In this section we assume that we are given a
conditional distribution function \(G\) such that
\(G(t \, \vert \, A_k,\bm{X}) = P(C > t \, \vert \, A_k,\bm{X})\).
Based on \(G\), we define the inverse probability weighted outcome:
\begin{align}
  \tilde{Y} := \frac{ \1 \lbrace \tilde{T} \le t_0, \tilde{\Delta} = 1\rbrace}{{G}(\tilde{T}\!-
  \, \vert \, A_k, \bm{X})}.
  \label{eq:defi:Y1:Y2}
\end{align}
For this outcome, we show in Section \ref{sec:identifiable:ipw} below
that
\begin{align*}
  &\theta_{\mathrm{crude}}
    =
    \EE\big[    \EE[ \tilde{Y}  \, \vert \, \bm{X}=\bm{x}, A_k=1]-    \EE[ \tilde{Y}  \, \vert \, \bm{X}=\bm{x}, A_k=0 ] \big]. 
\end{align*}
The idea is that we can apply GRFs directly to our weighted outcome
\(\tilde{Y}\). This provides an estimator
\(\hat{\theta}_{\mathrm{crude}}(\bm{x})\) for the conditional effect
\(\theta_{\mathrm{crude}}(\bm{x})=P( T^1 \le t_0, \Delta^1=1 \, \vert
\, \bm{X}=\bm{x}) - P( T^0 \le t_0, \Delta^0=1 \, \vert \,
\bm{X}=\bm{x})\) and thereby an estimator for the corresponding
average effect
\( \hat{\bar{\theta}}_{\mathrm{crude}} = \frac{1}{n} \sum_{i=1}^n
\hat{\theta}_{\mathrm{crude}} (\bm{X}_i)\).  This leads to the
following two-step approach:
\begin{itemize}[align=left,labelsep=0.4pt]
\item[\textbf{Step 1.}\quad]{The conditional distribution function
    \(G\) is estimated based on the full dataset and is used to
    construct the weighted outcome \(\tilde{Y}\) as defined by
    Equation \eqref{eq:defi:Y1:Y2}. }\vspace{0.2cm}
\item[\textbf{Step 2.}\quad]{A generalized random forest is applied
    with \(\tilde{Y}\) as outcome, yielding estimates
    \(\hat{\theta}_{\mathrm{crude}}(\bm{x})\),
    \(\bm{x}\in\mathcal{X}\), and the ATE is then estimated simply by
    averaging.  }
\end{itemize}
An equivalent two-step approach is utilized to estimate the effect on
net probabilities. We note that this requires, in addition to an
estimator for the conditional distribution \(G\), an estimator for the
conditional distribution function \(G_2\) such that
\(G_2(t \, \vert \, A_k, \bm{X}) = P(T^{2,a} > t \, \vert \, A_k,
\bm{X} ) \), and construction of the inverse probability weighted
outcome
\begin{align}
  \tilde{Y}' := \frac{ \1 \lbrace \tilde{T} \le t_0, \tilde{\Delta} = 1\rbrace}{{G}(\tilde{T}\! - 
  \, \vert \, A_k, \bm{X}){G}_2(\tilde{T}\! - 
  \, \vert \, A_k, \bm{X})}.
  \label{eq:defi:Y1:Y2:2}
\end{align}
Thus, to construct the weights, we need to model the survival
functions of both the latent time to a competing risk event and the
censoring time.

\subsection{Identifiability by inverse probability weighting}
\label{sec:identifiable:ipw}

The causal assumptions (see Section \ref{sec:identifiability:crude})
allow us to link the distribution of the counterfactual variables to
the observed data distribution. Since,
\begin{align*}
  &\EE\big[ \tilde{Y}  \, \big\vert \, \bm{X}, A_k\big]
    = \EE \left[ \frac{ \1 \lbrace \tilde{T} \le t_0, \tilde{\Delta} = 1\rbrace}{{G}(\tilde{T}\! - 
    \, \vert \, A_k, \bm{X})} \,\bigg\vert\, \bm{X}, A_k \right] 
    =\EE \big[ \1 \lbrace {T} \le t_0, \Delta =1 \rbrace
    \, \vert \, A_k, \bm{X} \big] , 
\end{align*}
it follows that,
\begin{align*}
  \bar{\theta}_{\mathrm{crude}}
  &  =  \EE \big[ P( T^1 \le t_0, \Delta^1=1  \, \vert \, \bm{X}=\bm{x}) -
    P(  T^0  \le t_0, \Delta^0=1  \, \vert \, \bm{X}=\bm{x}) \big] \\
  &  = \EE \big[ P( T \le t_0, \Delta=1  \, \vert \, \bm{X}=\bm{x}, A_k=1) -
    P(  T  \le t_0, \Delta=1  \, \vert \, \bm{X}=\bm{x}, A_k=0) \big]\\ 
  &  = \EE\big[\EE[ \tilde{Y} \, \vert \, \bm{X}=\bm{x}, A_k=1 ] -
    \EE[ \tilde{Y} \, \vert \, \bm{X}=\bm{x} , A_k=0 ]\big]. 
\end{align*}
Similarly, we identify \({\theta}_{\mathrm{net}}(\bm{x})\). More
details can be found in the supplementary material (Appendix B).

\subsection{Estimation of inverse probability weights}
\label{sec:weight:estimation}

To implement our two-step approach, we need consistent estimators for
the nuisance parameters \(G\) and \(G_2\) on \([0,t_0]\).  We here
describe an approach based on the reverse Kaplan-Meier estimator
stratified on a subset of categorical covariates
\(\bm{Z} \subset \lbrace A_k, \bm{X}\rbrace\). With this method, we
estimate the censoring survival distribution function \(G\),
conditional on \(\bm{Z}\), as follows
\begin{align*}
  &  \hat{G} (t \, \vert \, \bm{z}) =
    \prod_{t_k \le t} \left( 1- \frac{
    \sum_{i=1}^n \1 \lbrace T_i = t_k, \Delta_i =0 ,
    \bm{Z}_i =\bm{z} \rbrace
    }{
    \sum_{i=1}^n \big(\1 \lbrace T_i \ge t_k\rbrace - \1 \lbrace T_i = t_k,
    \Delta_i 
    >0 ,\bm{Z}_i =\bm{z}\rbrace \big)  
    } \right).
\end{align*}
Ties in the event times are handled with the usual convention that the
event of interest happens before competing events and censoring
events. Similarly, we estimate \(G_2\) with Kaplan-Meier estimator for
the competing event time conditional on \( \bm{Z}\),
\begin{align*}
  & \hat{G}_2 (t \, \vert \, \bm{z})= \prod_{t_k \le t} \left( 1- \frac{
    \sum_{i=1}^n  \1 \lbrace T_i = t_k, \Delta_i=2, 
    \bm{Z}_i=\bm{z}\rbrace
    }{
    \sum_{i=1}^n (\1 \lbrace T_i \ge t_k\rbrace - \1 \lbrace T_i = t_k, \Delta_i \neq 2
    ,
    \bm{Z}_i=\bm{z}\rbrace) } \right).
\end{align*}
Under the working assumption that
\( G(t \, \vert \, A_k, \bm{X} ) = P(C > t \, \vert \, A_k, \bm{X}) =
P(C > t \,\vert \, \bm{Z} )\), standard arguments \citep[][]{abgk}
lead to
\( \hat{G}(t|\bm{Z})\to G(t|\bm{Z})\,\, \text{ a.s. as } n\to\infty\)
for all \(t\le t_0\), and likewise for \(\hat{G}_2(t|\bm{Z})\).
However, violation of the working assumption may lead to asymptotic
bias in Step 1 of our two-step approach which may also lead to bias in
the ranking of the treatment variables.  We note that these working
assumptions are appropriate in our illustrative data example (Section
\ref{sec:data:application}), whereas other settings may require a
different approach; in Section \ref{sec:discussion}, we discuss the
bias-variance trade-off and how one may relax the working assumptions.

\section{Variable importance}
\label{sec:variable:importance}

Suppose we have a list of treatments, \(A_1, A_2, \ldots, A_K\),
\(K\in\N\), that we would like to rank according to their effect on a
time-to-event outcome.  Specifically for the purpose of ranking, we
continue our discussion from Section \ref{sec:competing:risks:problem}
to distinguish between crude and net probabilities. The problem with
crude probabilities is that they reflect a mixture of effects on the
hazard rate of the event of interest and effects on the hazard rate of
the competing risks.  Net effects, on the other hand, are defined in a
hypothetical world where all competing causes are eliminated and allow
us to study the effect of a particular drug in a way that is
independent of the effect that this drug may have on the hazard rate
of the competing events.  We argue that, for the purpose of drug
discovery, it may be desirable to restrict the search to drugs that
have net effects.
 
To obtain a ranking of the treatments, we apply the two-step approach
of Section \ref{sec:grf:with:pseudo} which yields estimates
\(\hat{\bar{\theta}}_{\mathrm{net},k}\) for the treatment effects on
the net probability scale for all drugs \(A_k\), \(k=1,
\ldots,K\). For comparison and illustration, we also compute estimates
\(\hat{\bar{\theta}}_{\mathrm{crude},k}\) for the treatment effect on
the crude probability scale.  A standard delta method argument using
the standard errors \(\hat{\sigma}_n(\bm{x})\) for the conditional
estimates \citep[as provided by][Theorem 5 and Section
6]{athey2019generalized} yields asymptotic normality of the forest
estimators \(\hat{\bar{\theta}}_{\mathrm{net},k}\),
\(\hat{\bar{\theta}}_{\mathrm{crude},k}\) for the average treatment
effects, based on which we construct confidence intervals.  Albeit the
asymptotic standard errors also contain a contribution from the
uncertainty of the weights constructed in Step 1 of our procedure,
these contributions are in our experience often very small in real
data applications. In our simulations and illustrative data analysis,
we only show confidence intervals which ignore the statistical
uncertainty due to Step 1. Despite these shortcomings, we note that in
our simulation studies (Section \ref{sec:simulation:strategy}) the
coverage of the confidence intervals lies nicely around 95\%.

\section{Simulation study}
\label{sec:simulation:strategy}

To evaluate the performance of our proposed methodology, and as a
proof of concept, we test our algorithm on simulated
data. \color{black} Our simulations further illustrate the difference
between treatment effects on the crude and net probability scales.
\color{black} We here explain the design of the simulations. Further
details in the form of \texttt{R}-code can be found on github, see
Section \ref{sec:supp:material}.

We start by simulating covariates, \(\bm{X}=(X_1,\ldots,X_6)\). We let
\(X_1, X_4, X_5, X_6\) be uniformly distributed on the unit interval
\((0,1)\), \(X_2\) be categorical with three ordered categories, and
\(X_4\) be categorical with four ordered categories. We consider a
setting where we compare \(K=10\) treatment variables drawn from
Bernoulli distributions that are all dependent on one of the
covariates,
\( \EE [A_k \, \vert \, \bm{X}] = \mathrm{expit}(\beta_0^k + \beta^k_1
X_{l_k}) \), with \(l_k\in \lbrace 1, \ldots, 6\rbrace\). Given
treatments and covariates, three latent event times \(T^1,T^2,C\) are
simulated according to Weibull distributions. The Weibull distribution
of the latent censoring time is specified independently of covariate
and treatment variables. The Weibull distribution of the latent time
to the event of interest is specified with a shape parameter dependent
on \(X_1\), \(X_3\) and \(A_1\). The Weibull distribution of the
latent competing event time is specified with a shape parameter
dependent on \(X_1\), \(X_2\) and \(A_2\).  Our simulation design is
summarized in Table \ref{table:simulations}.

\begin{table}[ht]
\centering
\begin{tabular}{llll}
  \midrule
  Event of interest: & \(T^1\sim A_1+X_1+X_3\) \\ 
  Competing event:  & \(T^2\sim A_2+X_1+X_2 \) \\
  Censoring: & \(C\sim 1\) \\ 
  \bottomrule
\end{tabular}
\caption{Summary of simulation design.  }
\label{table:simulations}
\end{table}

The parameters \(\bar{\theta}_{\mathrm{net}}\) and
\(\bar{\theta}_{\mathrm{crude}}\) are defined with time horizon
\(t_0=0.5\).  Generally, we say that a treatment \(A_k\) has a
protective effect on the net (crude) probability scale if
\(\bar{\theta}_{\mathrm{net},A_k}<0\)
(\(\bar{\theta}_{\mathrm{crude},A_k}<0\)), a harmful effect if
\(\bar{\theta}_{\mathrm{net},A_k}>0\)
(\(\bar{\theta}_{\mathrm{crude},A_k}>0\)), and a neutral effect if
\(\bar{\theta}_{\mathrm{net},A_k}=0\)
(\(\bar{\theta}_{\mathrm{crude},A_k}=0\)). Throughout this section, we
focus on three of the treatment variables: \(A_1\) that has a direct
effect on the event of interest, \(A_2\) that has a effect only on the
competing event, and \(A_3\) that has no effect at all. The values of
\(\bar{\theta}_{\mathrm{net},A_k},\bar{\theta}_{\mathrm{crude},A_k}\),
\(k=1,2,3\), can be found in Table \ref{table:effects}. Note that
\(A_2\) has no net effect but a protective crude effect since \(A_2\)
increases the rate of the competing event.

\begin{table}[ht]
\centering
\begin{tabular}{llll}
  \midrule
  Effects on net probabilities: & \(\bar{\theta}_{\mathrm{net},A_1}=-0.113\)
  &
    \(\bar{\theta}_{\mathrm{net},A_2} =0\) & \( \bar{\theta}_{\mathrm{net},A_3}=0\)\\ 
  Effects on crude probabilities: & \(\bar{\theta}_{\mathrm{crude},A_1}=-0.083\)
  & \(\bar{\theta}_{\mathrm{crude},A_2} = -0.047\)
    & \(\bar{\theta}_{\mathrm{crude},A_3}=0\) 
  \\
  \bottomrule
\end{tabular}
\caption{Values of
  \(\bar{\theta}_{\mathrm{net},A_k},\bar{\theta}_{\mathrm{crude},A_k}\),
  \(k=1,2,3\), for the simulation study.  }
\label{table:effects}
\end{table}

Our aim is to show that weighting yields unbiased estimation of the
ATEs and further to explore the effect of confounding and sample
size. Our simulations consist of the following two parts:
\begin{itemize}[align=left,labelsep=0.4pt]
\item[\textit{1. Effect estimation and coverage.}\quad]{We simulate
    \(M=500\) datasets with sample size \(n=500\) from the
    data-generating distribution. We look at effect estimates and
    coverage of the confidence intervals based on the standard error
    estimates provided by the forest. } \vspace{0.2cm}
\item[\textit{2. Ranking effectiveness.}\quad]{For sample sizes
    \(n \in \lbrace 100, 200,\) \( 500, 1000, 1500, 2000\rbrace\), we
    simulate \(M=500\) datasets from the data-generating
    distribution. For each dataset, we use our algorithm to estimate
    the variable importance of the treatments, in form of estimates
    \(\hat{\bar{\theta}}^m_{\mathrm{net},A_k}\), and
    \(\hat{\bar{\theta}}^m_{\mathrm{crude},A_k}\) for
    \(k=1,\ldots, 10\) and \(m=1,\ldots, M\). For \(k=1,\ldots, 10\),
    we define
    \begin{align}
      \mathcal{R}^M_{\mathrm{net}}(A_k) &:= \frac{1}{M}\sum_{m=1}^M
      \prod_{k' \neq k} \1 \lbrace
                             \hat{\bar{\theta}}^m_{\mathrm{net},A_k} \le \hat{\bar{\theta}}^m_{\mathrm{net},A_{k'}} \rbrace,
                                          \label{eq:R1}\\
      \mathcal{R}^M_{\mathrm{crude}}(A_k) &:= 
                                            \frac{1}{M}\sum_{m=1}^M
                                            \prod_{k' \neq k} \1 \lbrace
                                            \hat{\bar{\theta}}^m_{\mathrm{crude},A_k}
                                            \le \hat{\bar{\theta}}^m_{\mathrm{crude},A_{k'}} \rbrace,
       \label{eq:R1:R2}
    \end{align}
    as the fraction of simulation repetitions (out of \(M=500\)) where
    the treatment variable \(A_k\) is ranked ``most important'' among
    \(A_1,\ldots,A_{10}\) in terms of the protective effect on net and
    crude probabilities, respectively.  We report the ability of our
    method to, for instance, detect treatment \(A_1\) as the ``most
    important'' variable among \(A_1,\ldots, A_{10}\).  }
\end{itemize}
We consider three different adjustment schemes for the inverse
probability weight estimation:
\begin{itemize}[align=left,labelsep=0.4pt]
\item[{(a)} \,] Weight estimators \(\hat{G},\hat{G}_2\) that are
  adjusted for \(A_2\), \(X_1\) and \(X_2\), i.e.,
  \(\bm{Z} = \lbrace A_2, X_1, X_2 \rbrace\).
\item[{(b)} \,] Weight estimators \(\hat{G},\hat{G}_2\) that are
  adjusted only for \(A_2\), i.e., \(\bm{Z} = \lbrace A_2 \rbrace\).
\item[{(c)} \,] Weight estimators \(\hat{G},\hat{G}_2\) that are
  unadjusted, i.e., \(\bm{Z}=\lbrace 1\rbrace\).
\end{itemize}\color{black}\color{black}
The weight estimators are constructed outside the forest in Step 1 of
our two-step procedure as described in Section
\ref{sec:variable:importance}. Based on the weights, a separate (GRF)
forest is applied for each treatment variable \(A_k\) to estimate
\(\bar{\theta}_{\mathrm{net},A_k}\) and
\(\bar{\theta}_{\mathrm{crude},A_k}\), for \(k=1,\ldots, 10\).

\subsection{Simulation results}
\label{sec:simulation:application}

\subsubsection{Effect estimation and coverage}

Figure \ref{fig:plot:ranking} shows mean estimates across \(M=500\)
simulated datasets using adjustment schemes (a)--(c) for estimation of
inverse probability weights. Using adjustment scheme (a), treatment
\(A_1\) is correctly shown to have a protective effect both on the
scale of net probabilities and on the scale of crude
probabilities. Furthermore, treatment \(A_2\) is correctly shown to
have a protective effect on the crude probabilities and no effect on
the net probabilities and treatment \(A_3\) is correctly shown to have
no effect on both scales. Confidence intervals all have a coverage
around \(95 \%\) despite the fact that the standard errors do not take
the uncertainty of the weight estimation into account.  Using
adjustment scheme (b) the weights are only adjusted for \(A_2\), but
Figure \ref{fig:plot:ranking} shows that we still achieve \(95\%\)
coverage with our confidence intervals.  A comparison of the results
for adjustment schemes (b) and (c) in Figure \ref{fig:plot:ranking},
on the other hand, reveals that it is crucial to include treatment
\(A_2\) in \(\bm{Z}\) in the weight estimation for estimating
\(\bar{\theta}_{\mathrm{net}}\): Adjustment scheme (c) uses unadjusted
estimators for the inverse probability weights leading to an incorrect
conclusion of a protective effect of treatment \(A_2\) on the net
probabilities.

Estimation of \(\bar{\theta}_{\mathrm{crude}}\) is hardly affected
across the weighting schemes (a)--(c) since the censoring times were
generated independent of all treatment and covariate variables. Of
course, we can produce biased results for
\(\bar{\theta}_{\mathrm{crude}}\) with the unadjusted weighting scheme
if we let the censoring mechanism depend on treatment variables and
covariates.

\begin{figure}[!h]   
  \centering \makebox[\textwidth][c]{\includegraphics[width=1.0\textwidth,angle=0]
  {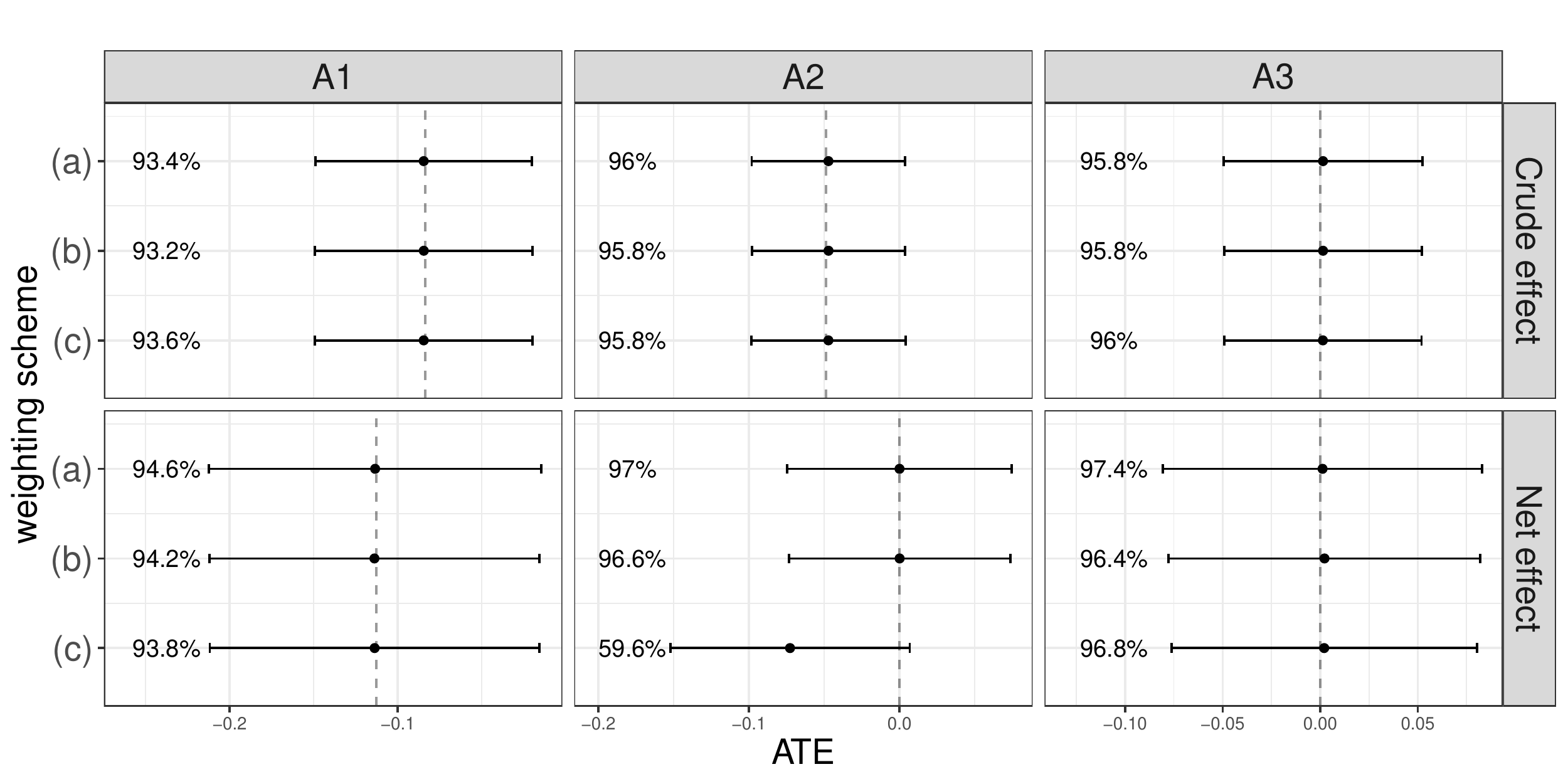}}
\caption{Results of the simulation studies. Shown are the results from
  estimation of \(\bar{\theta}_{\mathrm{net},A_k}\) and
  \(\bar{\theta}_{\mathrm{crude},A_k}\), \(k=1,2,3\), across \(M=500\)
  repetitions (all with sample size \(n=500\)). The true values are
  marked by the dashed gray lines. Note that \(A_2\) has an effect on
  the difference in crude probabilities (true effect
  \(\bar{\theta}_{\mathrm{crude},A_2} = -0.047\)) through the effect
  on the competing risk event whereas it has no effect on the
  difference in net probabilities (true effect
  \(\bar{\theta}_{\mathrm{net},A_2}=0\)).  The right column shows the
  coverage, i.e, the fraction of simulations where the confidence
  interval constructed based on the forest estimate of the standard
  error contains the true value. In weighting schemes (a) and (b), we
  used weights that were adjusted for
  \(\bm{Z}=\lbrace A_2, X_1, X_2\rbrace\) and
  \(\bm{Z}=\lbrace A_2\rbrace\), respectively, both resulting in
  unbiased estimators. In weighting scheme (c) we used unadjusted
  weights (\(\bm{Z}=\lbrace 1\rbrace\)), inducing severe bias in the
  estimate of \(\bar{\theta}_{\mathrm{net},A_2}\). We do not show the
  results for \(A_4, \ldots, A_{10}\) as these are similar to those
  for \(A_3\).  }
  \label{fig:plot:ranking}   
\end{figure}

\subsubsection{Ranking effectiveness}

Figure \ref{fig:plot:ranking} shows the fractions
\(\mathcal{R}^M_{\mathrm{net}}(A_1)\) and
\(\mathcal{R}^M_{\mathrm{crude}}(A_1)\) defined in Equations
\eqref{eq:R1} and \eqref{eq:R1:R2} across different sample sizes. We
show only the results from using adjustment scheme (b) and adjustment
scheme (c) for estimating the inverse probability weights, as the
results for weighting scheme (a) and (b) are
similar.

Figure \ref{fig:plot:ranking} shows
\(\mathcal{R}^M_{\mathrm{net}}(A_k)\) and
\(\mathcal{R}^M_{\mathrm{crude}}(A_k)\) for each of the three
treatment variables \(A_1\), \(A_2\) and \(A_3\). Recall that these
are the fractions of simulation repetitions where \(A_k\) is ranked
most important among \(A_1,\ldots, A_{10}\) in terms of their effects
on the net and crude probabilities, respectively. We would like
\(\mathcal{R}^M_{\mathrm{net}}(A_1),
\mathcal{R}^M_{\mathrm{crude}}(A_1)\) to be close to one, and
\(\mathcal{R}^M_{\mathrm{net}}(A_2),
\mathcal{R}^M_{\mathrm{crude}}(A_2),
\mathcal{R}^M_{\mathrm{net}}(A_3),\)
\( \mathcal{R}^M_{\mathrm{crude}}(A_3)\) to be close to zero. We
further expect \(\mathcal{R}^M_{\mathrm{crude}}(A_2)\) to be larger
than \(\mathcal{R}^M_{\mathrm{net}}(A_2)\), due to the effect of
\(A_2\) on the competing risk event.

Figure \ref{fig:plot:ranking} shows that both
\(\mathcal{R}^M_{\mathrm{net}}(A_1) \) and
\(\mathcal{R}^M_{\mathrm{crude}}(A_1) \) approach one as the sample
size \(n\) increases: The larger the sample size, the more certain we
are to detect the important variable \(A_1\). On the other hand, it
also shows that \(\mathcal{R}^M_{\mathrm{net}}(A_1)\) and
\(\mathcal{R}^M_{\mathrm{crude}}(A_1) \) are both rather small for
\(n=100\) and \(n=200\). Evidently, we need a certain sample size to
be able to detect important variables with high probability. Across
all sample sizes we have that
\(\mathcal{R}^M_{\mathrm{net}}(A_3),
\mathcal{R}^M_{\mathrm{crude}}(A_3) \) are both very small, consistent
with the fact that \(A_3\) has no effect at all
(\(\bar{\theta}_{\mathrm{net},A_3}=\bar{\theta}_{\mathrm{crude},A_3}=0\)). The
same is seen for \(\mathcal{R}^M_{\mathrm{net}}(A_2) \), except in the
scenario where we fail to adjust for \(A_2\) in the estimation of
inverse probability weights. At last we note that
\(\mathcal{R}^M_{\mathrm{crude}}(A_2) \) is overall larger than
\(\mathcal{R}^M_{\mathrm{net}}(A_2) \), as we would expect.

\begin{figure}[!h]   
  \centering \makebox[\textwidth][c]{\includegraphics[width=1.0\textwidth,angle=0]
  {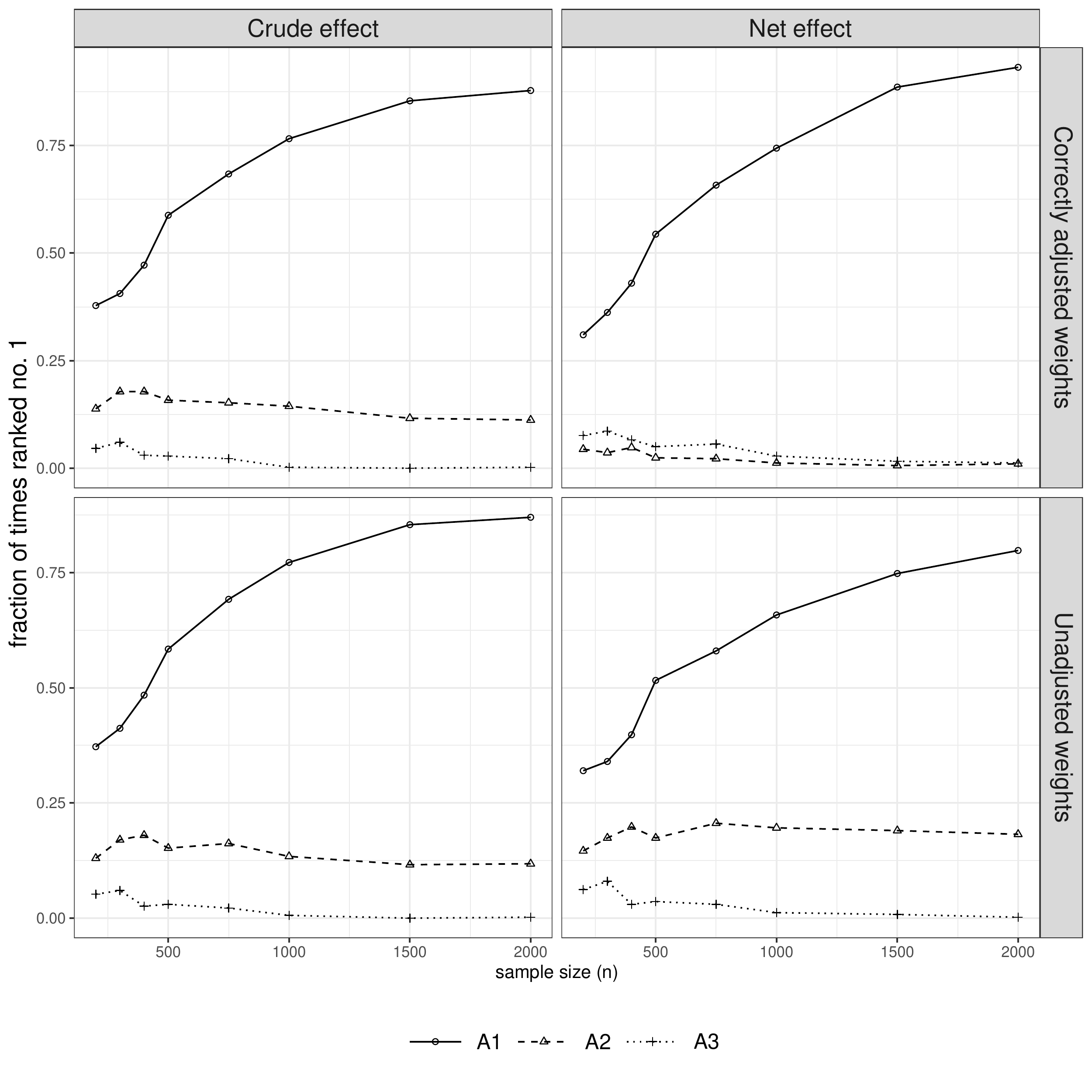}}
\caption{Results of the simulation studies. Shown are the fraction of
  times that each the three treatment variables \(A_1, A_2, A_3\) was
  ranked most important (across \(M=500\) simulation repetitions) in
  terms of either the effect on the difference in net probabilities
  (\(\bar{\theta}_{\mathrm{net}}\)) or the effect on the difference in
  crude probabilities (\(\bar{\theta}_{\mathrm{crude}}\)). In the left
  plot, estimation of the inverse probability weights were adjusted
  for \(A_2\) (weighting scheme (b)).  In the right plot, we used
  unadjusted estimators (weighting scheme (c)) for the inverse
  probability weights. }
  \label{fig:plot:ranking}   
\end{figure}

\section{Registry study}
\label{sec:data:application}

We apply our method to our motivating example in which it is of
interest to study whether the use of any particular drug decreases the
risk of relapse of depression resulting in psychiatric
hospitalization. We here report estimates of effects on the net
probabilities as well as those on the crude probabilities. Our aim is
to discover new active substances; for this purpose, net probabilities
will allow us to rank drugs according to their direct effect on
depression, isolating this effect from what effect that drug may have
on competing events.

\begin{figure}[!h]   
  \centering \makebox[\textwidth][c]{\includegraphics[width=1.0\textwidth,angle=0]
  {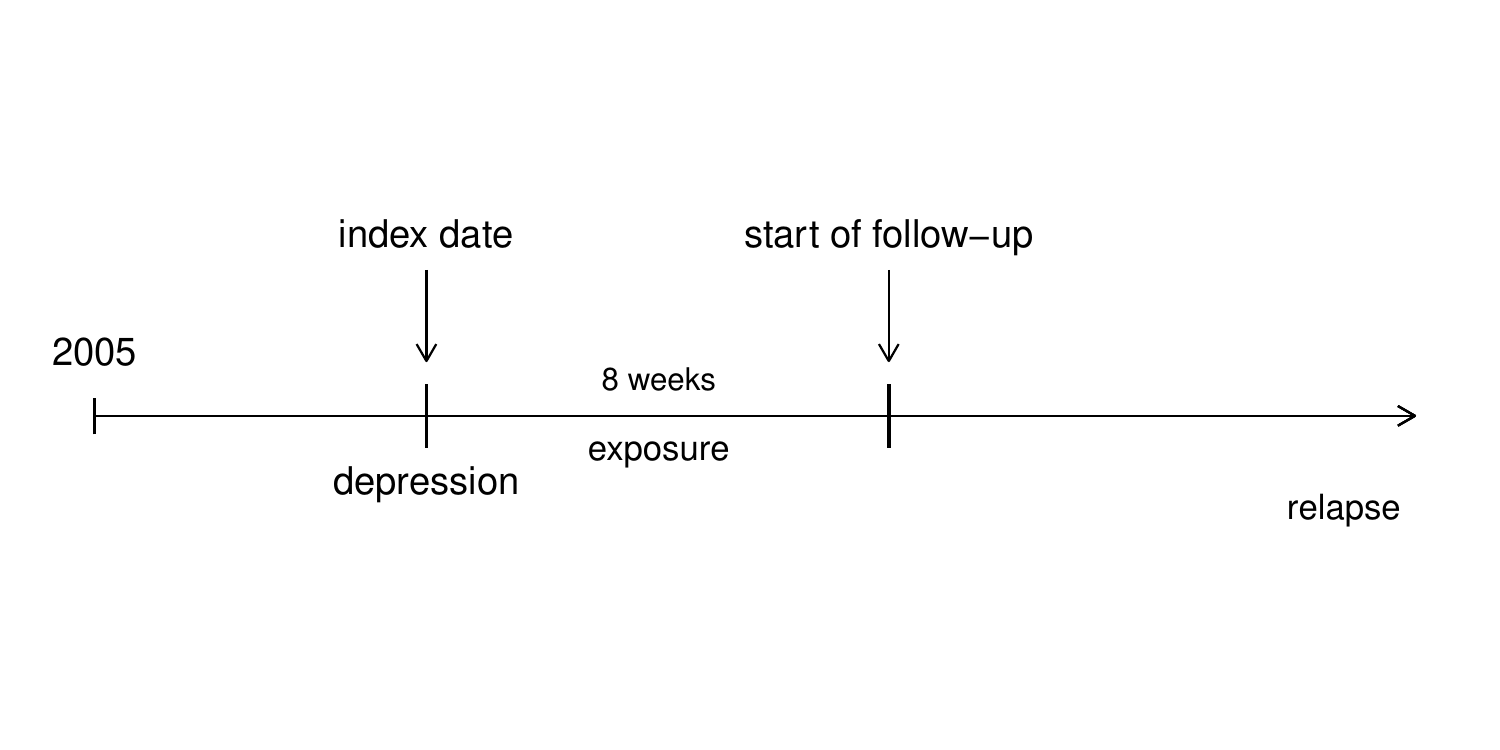}}
\caption{Illustration of our study design. The date of first contact
  with depression is defined as the index date. Patients with a
  psychiatric hospitalization in the eight weeks window following the
  index date are excluded.  ATC drug codes are grouped after their
  first three digits to define binary exposure variables with the
  value \(1\) if there was at least one prescribed purchase within the ATC
  group in the eight weeks window.  Information on comorbidity is
  collected during a ten year period before the index date and
  included as covariates in the analysis, along with sex and age at
  the index date.}
  \label{fig:design:grf}   
\end{figure}

The data we work with are obtained by linking Danish population-based
registers that contain data on all prescribed medical purchases at
pharmacies since 1995 and data on all patients treated at hospitals
since 1977. A total of 78,700 patients were included who all had a
first-time admission with depression after 2005.  Figure
\ref{fig:design:grf} illustrates our design.  The date of first
contact with depression was defined as the index date. Patients with a
psychiatric hospitalization in the eight weeks window following the
index date were excluded.  ATC drug codes were grouped after their
first three digits to define binary exposure variables \(A_k\) with
the value 1 if there was at least one prescribed purchase within the
ATC group in the eight weeks window. Information on comorbidity was
collected during a ten year period before the index date and included
as covariates in the analysis, along with sex and age at the index
date.  Subjects were followed for five years from the end of the
exposure window until depression relapse (\(\Delta=1\)), a competing
event (\(\Delta=2\)), or loss to follow-up (\(\Delta=0\)). Summary
statistics on comorbidities, exposure and number of events can be
found in the supplementary material (Appendix D).

To estimate the treatment effect of each considered drug group \(A_k\)
on the net and crude probabilities,
\(\bar{\theta}_{\mathrm{net}, A_k}\) and
\(\bar{\theta}_{\mathrm{crude}, A_k}\), the inverse probability
weights were adjusted for sex, age group and the treatment \(A_k\)
itself. In the forest we used \(B=200\) trees, and we included sex,
age group and all comorbidities as covariates.

\subsection{Results}
\label{sec:data:results}

\begin{figure}[!h]   
  \centering \makebox[\textwidth][c]{\includegraphics[width=1.2\textwidth,angle=0]
  {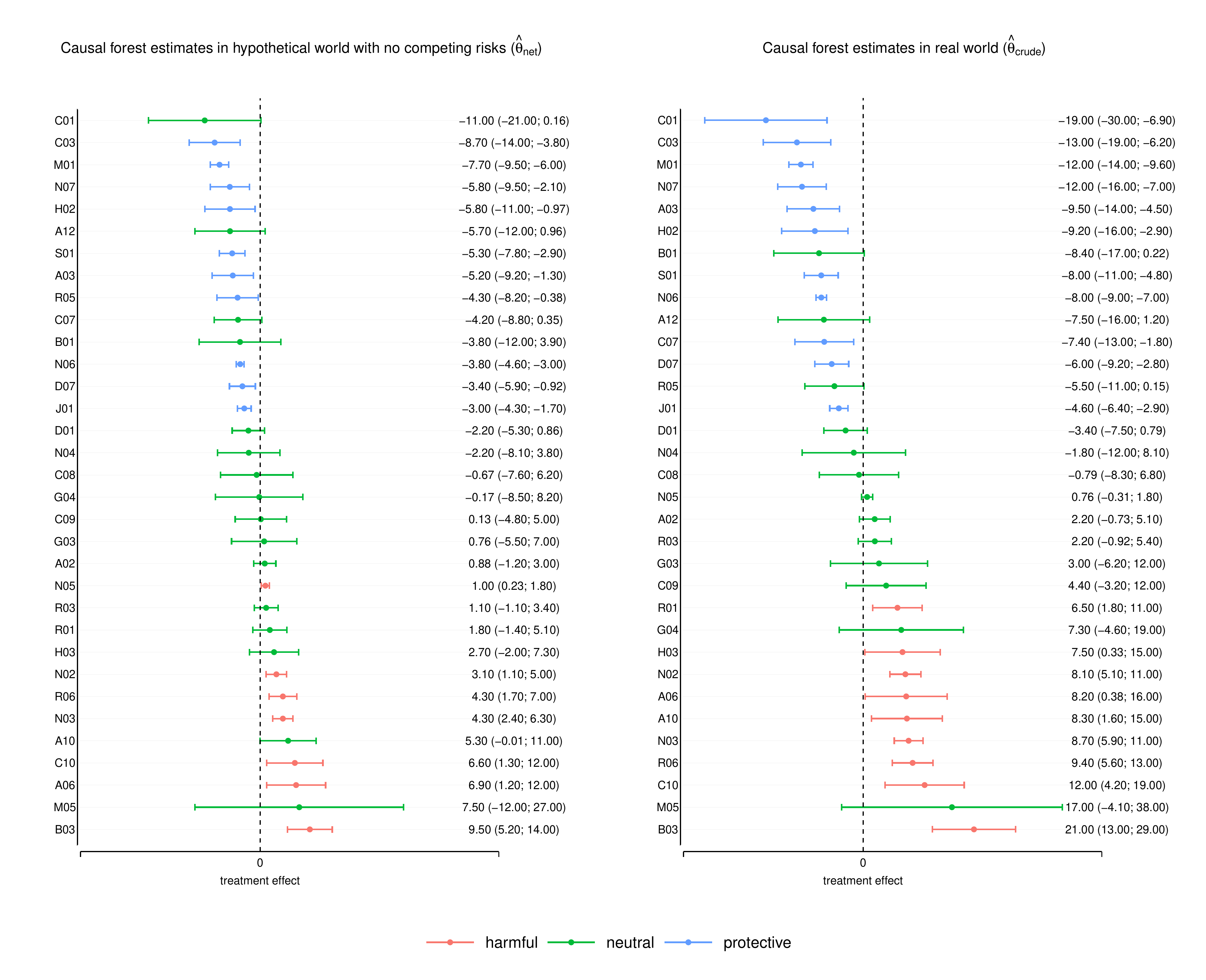}}
\caption{\textit{Left:} Causal forest estimates of
  \(\bar{\theta}_{\mathrm{net}}\) (using adjusted weights to construct
  weighted outcomes). \textit{Right:} Causal forest estimates of
  \(\bar{\theta}_{\mathrm{crude}}\) (using adjusted weights to
  construct weighted outcomes). For each ATC group (marked on the
  \(x\)-axis) the plot shows the estimates and the estimated confidence
  intervals (numbers written on the right). The colors indicate the
  direction of the effect.} 
  \label{fig:plot2} 
\end{figure}

Figure \ref{fig:plot2} shows the causal forest estimates of the effect
on net probabilities, \(\bar{\theta}_{\mathrm{net}}\), and of the
effect on crude probabilities, \(\bar{\theta}_{\mathrm{crude}}\), for
each drug group. We distinguish between a protective effect (if the
upper confidence limit is below zero), a harmful effect (if the lower
confidence limit is above zero), and a neutral effect (if zero is
contained in the confidence interval). The size of the estimates
allows us to rank the treatment groups according to their effect on
relapse with depression. As we saw in the simulation study, there can
be a substantial difference between \(\bar{\theta}_{\mathrm{net}}\)
and \(\bar{\theta}_{\mathrm{crude}}\).  Here we see in Figure
\ref{fig:plot2}, as well, that the estimates of the two parameters
lead to slightly differing conclusions. Consider, for example, the
drug group `A12' (mineral supplements). This drug group is ranked
higher in terms of net probabilities than in terms of crude
probabilities (although the effect remains insignificant in both
cases).  On the other end of the spectrum, some drug groups are deemed
harmful in terms of their effect on crude probabilities and neutral in
terms of their effect on net probabilities: `A10' (antidiabetics) and
`C10' (lipid modifying agents). Recall that net effects, if we believe
in the assumptions required to go from a crude to a net interpretation
(Section \ref{sec:identifiability:crude}), allow us to rank drugs
according to their direct effect on the depression relapse without
interference from indirect effects on the competing events. Thus, we
can avoid pitfalls like reporting, as we saw in our simulation study,
a large treatment effect simply if that treatment increases the rate
of a competing event, or, as we saw in Example
\ref{ex:hazard:vs:risk}, concluding smaller effects of treatments that
also have protective effects on the rate of competing
events.

\section{Discussion}
\label{sec:discussion}

In this paper we have considered average treatment effect estimation
for the purpose of ranking treatments according to their effect on a
specific time-to-event outcome of interest. We have implemented a
data-adaptive estimation method based on generalized random forests,
where inverse probability weights are constructed to make the forest
implementation directly applicable to the time-to-event setting.  Our
method makes no parametric model restrictions and really benefits from
the flexibility of the generalized random forest which adaptively
adjusts the propensity of treatment for covariates. This altogether
makes it highly applicable to drug discovery studies with many
candidate drug treatments and not much prior subject matter knowledge.
As an illustration, we have considered a particular application where
it was of interest to rank a list of treatments according to their
effect on depression.

To handle competing risks, we have discussed the use of two different
average treatment effect parameters in the presence of competing
risks, defined in terms of net and crude probabilities, respectively,
with different interpretations. Particularly, net probabilities allow
us to make inference for treatment directly on the outcome of
interest, irrespective of that treatment's effect on competing
risks. We argue for the utility of net probabilities when looking for
new active substances as part of a drug discovery study, but
emphasize, in accordance with earlier criticism, that they are not
sensible interpreting the size of the effect, e.g., when counseling a
patient.  Crude probabilities should always be considered if interest
is in the real world and the aim is to predict for a given
patient. The methods proposed recently by \cite{stensrud2020separable}
provide an alternative route for isolating direct effects on the event
of interest, but their methods require other untestable assumptions on
the biological nature of the treatment mechanism.

A weakness of our presented analysis is the use of the Kaplan-Meier
method for constructing the inverse probability weights. This may work
in large scale registry data where most variables are categorical and
a large amount of data are available to estimate the weights
separately in all strata defined by the covariates. However, in other
applications it may be necessary to allow that several continuous
covariates affect the distributions \(G\) and \(G_2\).  Semiparametric
theory tells us to use a flexible model and to include all covariates
that affect the event time to improve robustness and efficiency
\citep{vanRobins2003unified}. However, to achieve proper bias-variance
trade-off for the target parameter in the second step, a data-adaptive
method used for the weights must be undersmoothed.  Another idea is to
handle the weight estimation inside the forest in a one-step approach.
Indeed, we may improve upon the current setting by implementing the
splitting rule based on the efficient influence function
\citep{robins1992recovery,vanRobins2003unified}, extending the methods
of \cite{rytgaard2019grf} to the competing risks setting.  In future
work we follow this route and revise the implementation of GRFs to
adapt it to the event history analysis setting as proposed by
\cite{rytgaard2019grf}.

\section{Supplementary Material} 
\label{sec:supp:material}

\verb+R+ code is available on github
(\url{https://github.com/helenecharlotte/grfCausalSearch}). The
supplementary material consists of Appendices A--D.

\newpage\appendix

\section*{Appendix A}

\setcounter{section}{1}

We here detail the identifiability assumptions for the effect on net
probabilities and the effect on crude probabilities, respectively.

\subsection{Identifiability assumptions for the effect on net probabilities, \(\theta_{\mathrm{net}}(\bm{x})\)}
\label{sec:causal:assumptions:1}

Identification of \(\theta_{\mathrm{net}}(\bm{x})\) in terms of the
observed data distribution depends on three untestable causal
assumptions: Consistency, coarsening at random and positivity.

First, the assumption of consistency entails that the counterfactual
event time \(T^{1,a}\) corresponds to the observed event time for
those subjects who were actually uncensored, free of event type
\(j=2\) and were exposed to the treatment level
\(A_k=a\). Particularly, consistency provides the counterfactual
variables as follows:
\begin{align} 
  \tag{1a}\label{ass:1a} 
T = \min (T^{1,A_k}, T^{2,A_k}), \text{ and that, }
  {T}^{1,a}= {T}^{1,A_k} \text{ on the event that } A_k=a \text{ for }
  a=0,1. 
\end{align}
Here \(T^{2, A_k}\) is the uncensored counterfactual event time of
type \(j=2\) under the observed treatment.

The second assumption of coarsening at random is characterized as
follows. The full data we would have liked to observe are
\((\bm{X},T^{1,0},T^{1,1})\). These are not fully observed due to
censoring, the competing event and the treatment decision \(A_k\), and
we observe only the coarsened data
\((\bm{X}, {A_k},\tilde{T},\tilde{\Delta})\)
\citep{gill1997coarsening,vanRobins2003unified,tsiatis2007semiparametric}. To
identify \((\bm{X},T^{1,0},T^{1,1})\) from the data, we need
coarsening at random (CAR) \citep[][Section
1.2.3]{gill1997coarsening,vanRobins2003unified}, i.e., that the
coarsening mechanism only depends on the full data structure
\((\bm{X},T^{1,0},T^{1,1})\) through the observed data structure
\((\bm{X}, {A_k},\tilde{T},\tilde{\Delta})\).  Coarsening at random is
implied by the following conditional independence conditions:
\begin{align}
  \begin{split}
    & T^{1,a} \independent A_k \, \vert \, \bm{X}, \\
    & T^{1,A_k} \independent
    (C,T^{2,A_k}) \, \vert \, A_k, \bm{X} ,
  \end{split}\tag{1b}\label{ass:1b}
\end{align}
for \(a=0,1\), also refer to as ``no unmeasured confounding''.

The last assumption of positivity requires for the coarsening
mechanism that
\begin{align}
  \begin{split}
    & P(\min(C,T^{2,A_k}) \ge t_0 \, \vert \, A_k, \bm{X})\,
    (\pi_{k}(\bm{X}))^{A_k} (1-\pi_{k}(\bm{X}))^{1-A_k} >\eta>0,
  \end{split}
    \tag{1c}\label{ass:1c}
\end{align}
almost surely.

Under Assumptions \ref{ass:1a}, \ref{ass:1b} and \ref{ass:1c}, we can
link the distribution of the counterfactual variables to the observed
data distribution as follows:
\begin{align}
  \begin{split}
    & P( \tilde{T} \in dt, \tilde{\Delta}=1, A_k=a , \bm{X} \in d\bm{x})\\
    &\quad =
    P( \tilde{\Delta}=1 \, \vert \, T^{1,A_k}=t, A_k=a, \bm{X}=\bm{x}) P( T^{1,A_k} \in dt , A_k=a, \bm{X} \in d\bm{x}) \\
    &\quad = P( \min (C, T^{2,A_k}) \ge t \, \vert \, T^{1,A_k}=t,
    A_k=a, \bm{X}=\bm{x}) P( T^{1,A_k} \in dt \, \vert \, A_k=a,
    \bm{X}=\bm{x})
    \\
    &\qquad\qquad\qquad\qquad\quad\qquad\qquad\qquad
    \qquad\qquad\quad\qquad\qquad\qquad\quad\,\,\, P(A_k=a \, \vert \, \bm{X}=\bm{x}) P(\bm{X}\in d\bm{x}) \\
    &\quad = P( \min (C, T^{2,A_k}) \ge t \, \vert \, A_k=a,
    \bm{X}=\bm{x}) P( T^{1,a} \in dt \, \vert \, \bm{X}=\bm{x})
    P(A_k=a \, \vert \, \bm{X}=\bm{x}) P(\bm{X}\in d\bm{x}).
    \end{split}\label{eq:factorization:1}
\end{align}
Particularly, the first line of Assumption \ref{ass:1b} together
with the Assumption \ref{ass:1a} of consistency implies that
\begin{align*}
  P( T^{1,A_k} \in dt  \, \vert \,  A_k=a, \bm{X}\in d\bm{x})  
  &\overset{\ref{ass:1b}}{=}
  P( T^{1,a} \in dt  \, \vert \, A_k=a,  \bm{X}=\bm{x})\\
  &\overset{\ref{ass:1a}}{=}
  P( T^{1,a} \in dt  \, \vert \,  \bm{X}=\bm{x}),
\end{align*}
whereas the second line of Assumption \ref{ass:2a} yields that
\begin{align*}
  P( \min (C, T^{2,A_k}) \ge t \, \vert \, T=t,\Delta=1, A_k=a, \bm{X}=\bm{x})
  =  P(  \min (C, T^{2,A_k}) \ge t  \, \vert \,  A_k=a, \bm{X}=\bm{x}). 
\end{align*}
Assumption \ref{ass:1c} ensures that the right hand side of
\eqref{eq:factorization:2} is non-zero and well-defined.

\subsection{Identifiability assumptions for the effect on crude
  probabilities, \(\theta_{\mathrm{crude}}(\bm{x})\)}
\label{sec:causal:assumptions:2}

The assumptions needed to identify \(\theta_{\mathrm{crude}}(\bm{x})\)
are less restrictive than those needed for
\(\theta_{\mathrm{net}}(\bm{x})\) and correspond to the standard
setting for right-censored survival times.  The consistency assumption
for \(\theta_{\mathrm{crude}}(\bm{x})\) can be expressed as
\begin{align}
  \tag{2a}\label{ass:2a}
  T= T^a \text{ and } \Delta = \Delta^a \text{ on the event that } A=a, \, \text{for } a=0,1.
\end{align}
The full data we would have liked to observe are
\((\bm{X}, {T^0},{T^1},{\Delta^0},{\Delta^1})\), but we observe only
the coarsened data \((\bm{X}, {A_k},\tilde{T},\tilde{\Delta})\) due to
censoring \(C\) and treatment decision \(A_k\). The equivalent of
Assumption \ref{ass:1b},
\begin{align}
  \begin{split}
    & (T^a, \Delta^a) \independent A_k \, \vert \, \bm{X} , \quad \text{ for } a=0,1, \\
    & (T, \Delta) \independent C \, \vert \, A_k, \bm{X} ,
\end{split}\tag{2b}\label{ass:2b}
\end{align}
yields coarsening at random. We further make the positivity assumption
that,
\begin{align}
  &  P(C \ge t_0\, \vert \, A_k=a, \bm{X}) \,
    (\pi_{k}(\bm{X}))^{a} (1-\pi_{k}(\bm{X}))^{1-a} >\eta>0 ,\tag{2c}\label{ass:2c}
\end{align}
almost surely, for \(a=0,1\).

We can now express the observed data distribution as,
\begin{align}
  \begin{split}
  & P( \tilde{T} \in dt, \tilde{\Delta}=1, A_k=a , \bm{X} \in d\bm{x})\\
  &\quad =
    P( \tilde{\Delta} \ge 1  \, \vert \, T=t, \Delta=1, A_k=a, \bm{X}=\bm{x}) P( T \in dt , \Delta=1, A_k=a, \bm{X} \in d\bm{x}) \\
  &\quad =
    P( C \ge t \, \vert \, T=t,\Delta=1, A_k=a, \bm{X}=\bm{x}) P( T \in dt, \Delta=1  \, \vert \,  A_k=a, \bm{X}\in d\bm{x})
  \\
  &\qquad\qquad\qquad\qquad\quad\qquad\qquad\qquad
    \qquad\qquad\qquad\qquad\qquad P(A_k=a \, \vert \, \bm{X}=\bm{x}) P(\bm{X}\in d\bm{x}) \\
  &\quad =
    P( C \ge t \, \vert \,  A_k=a, \bm{X}=\bm{x}) P( T^{a} \in dt, \Delta^a=1  \, \vert \,  \bm{X}=\bm{x})
    P(A_k=a \, \vert \, \bm{X}=\bm{x}) P(\bm{X}\in d\bm{x}),
      \end{split} \label{eq:factorization:2}
\end{align}
relying on Assumptions \ref{ass:2a}, \ref{ass:2b} and
\ref{ass:2c}. Particularly, the first line of Assumption \ref{ass:2b}
together with the Assumption \ref{ass:2a} of consistency implies that
\begin{align*}
  P( T \in dt, \Delta=1  \, \vert \,  A_k=a, \bm{X}\in d\bm{x})  
  &\overset{\ref{ass:2b}}{=}
  P( T^{a} \in dt, \Delta^a=1  \, \vert \, A_k=a,  \bm{X}=\bm{x})\\
  &\overset{\ref{ass:2a}}{=}
  P( T^{a} \in dt, \Delta^a=1  \, \vert \,  \bm{X}=\bm{x}),
\end{align*}
whereas the second line of Assumption \ref{ass:2a} yields that
\begin{align*}
  P( C \ge t \, \vert \, T=t,\Delta=1, A_k=a, \bm{X}=\bm{x})
  =  P( C \ge t \, \vert \,  A_k=a, \bm{X}=\bm{x}). 
\end{align*}
Assumption \ref{ass:2c} ensures that the right hand side of
\eqref{eq:factorization:2} is non-zero and well-defined.

\section*{Appendix B}

\setcounter{section}{2}

\subsection{Weighted outcome for net probabilities}
Define the weighted outcome:
\begin{align*}
  \tilde{Y}' =
  \frac{ \1 \lbrace \tilde{T} \le t_0, \tilde{\Delta} = 1 \rbrace}{{G}'(\tilde{T} \!-  \, \vert \, A_k, \bm{X})}, 
\end{align*}
with weights given by
\begin{align*}
  {G}'(\tilde{T} \! - \, \vert \, A_k, \bm{X})
  = P (\min (T^{2,A_k}, C) \ge t  \, \vert \, A_k, \bm{X}). 
\end{align*}
For this weighted outcome we have that,
\begin{align}
  \begin{split}
    \theta_{\mathrm{net}}(\bm{x}) &= P( T^{1,1} \le t_0 \, \vert \,
    \bm{X}=\bm{x}) -
    P( T^{1,0} \le t_0  \, \vert \, \bm{X}=\bm{x} ) \\
    &= \EE\big[ \tilde{Y}' \, \big\vert \, \bm{X}=\bm{x}, A_k=1\big] -
    \EE\big[ \tilde{Y}' \, \big\vert \, \bm{X}=\bm{x}, A_k=0 \big].
                                \end{split}\label{eq:net:111}
\end{align}
This follows straightforwardly by the identification in, and just
after, Equation \eqref{eq:factorization:1}; indeed, we note that
\begin{align*}
\begin{split}
  \EE[ \tilde{Y}'\, \vert \, \bm{X}, A_k ] &= \EE\left[ \frac{ \1
      \lbrace \tilde{T}\le t_0, \tilde{\Delta}=1
      \rbrace}{{G}'(\tilde{T}\! - 
      \, \vert \, \bm{X}, A_k)} \, \bigg\vert\,  \bm{X}, A_k\right] \\
  &= \EE\left[ \EE\left[ \frac{ \1 \lbrace T^{1,A_k}\le t_0 \rbrace \1
        \lbrace \tilde{\Delta} =1\rbrace }{{G}'(\tilde{T}\! - 
        \, \vert \, \bm{X}, A_k)} \, \bigg\vert\, T^{1,A_k}, \bm{X}, A_k\right]\, \bigg\vert\,  \bm{X}, A_k\right] \\
  &= \EE \big[\1 \lbrace {T}^{1,A_k} \le t_0 \rbrace \,
  \vert\,\bm{X},A_k\big]\, \EE \left[ \frac{\EE [ \1 \lbrace
      \tilde{\Delta}=1\rbrace \, \vert \, T^{1,A_k},\bm{X}, A_k ] }{
      {G}'({T}^{1,A_k}\! -  \, \vert \, \bm{X}, A_k)} \right]
  \\
  &= \EE \big[\1 \lbrace {T}^{1,A_k} \le t_0 \rbrace \,
  \vert\,\bm{X},A_k\big]\, \EE \left[ \frac{ {G}'({T}^{1,A_k}\! -  \, \vert \, \bm{X}, A_k) }{
      {G}'({T}^{1,A_k}\! -  \, \vert \, \bm{X}, A_k)} \right]
  \\
  & = \EE[\1 \lbrace {T}^{1,A_k} \le t_0 \, \vert \, \bm{X}, A_k ],
\end{split}
\end{align*}
and
\begin{align*}
  \EE[ \1 \lbrace {T}^{1,A_k} \le t_0  \, \vert \,  \bm{X}, A_k=a ]
  &= \EE[\1 \lbrace {T}^{1,a} \le t_0 \rbrace\, \vert \,  \bm{X}, A_k=a ] 
    = \EE[\1 \lbrace {T}^{1,a} \le t_0 \rbrace\, \vert \,  \bm{X} ] ,
\end{align*}
for \(a=0,1\), which yields \eqref{eq:net:111}.

\subsection{Weighted outcome for crude probabilities}

For the weighted outcome,
\begin{align*}
  \tilde{Y}= \frac{ \1 \lbrace \tilde{T} \le t_0, \tilde{\Delta} = 1\rbrace}{{G}(\tilde{T}\! - 
  \, \vert \, A_k, \bm{X})},
\end{align*}
we have that,
\begin{align}
  \begin{split}
  \theta_{\mathrm{crude}}(\bm{x})& = P( T^1 \le t_0, \Delta^1=1  \, \vert \, \bm{X}=\bm{x}) -
                                   P(  T^0  \le t_0, \Delta^0=1  \, \vert \, \bm{X}=\bm{x}) \\
                                 & =
                                   \EE\big[ \tilde{Y}  \, \big\vert \, \bm{X}=\bm{x}, A_k=1\big] -
                                   \EE\big[ \tilde{Y}  \, \big\vert \, \bm{X}=\bm{x}, A_k=0 \big] .
                                   \end{split}\label{eq:crude:111}
\end{align}
This follows straightforwardly by the identification in, and just
after, Equation \eqref{eq:factorization:2}; indeed, we note that
\begin{align*}
  \EE[ \tilde{Y} \, \vert \, \bm{X}, A_k ]
  &= \EE\left[ \frac{ \1
    \lbrace \tilde{T}\le t_0, \tilde{\Delta}=1
    \rbrace}{{G}({\tilde{T}}\! - 
    \, \vert \, \bm{X}, A_k)} \, \bigg\vert\,  \bm{X}, A_k\right] \\
  &= \EE\left[ \EE\left[ \frac{ \1 \lbrace T\le t_0, \Delta=1 \rbrace
    \1\lbrace \tilde{\Delta} \ge 1 \rbrace}{{G}({T}\! - 
    \, \vert \, \bm{X}, A_k)} \, \bigg\vert\, T, \Delta, \bm{X}, A_k\right]\, \bigg\vert\,  \bm{X}, A_k\right] \\
  &= \EE \left[\1 \lbrace {T} \le t_0 , \Delta= 1\rbrace \, \frac{
    \EE\big[ \1 \lbrace \tilde{\Delta} \ge 1 \rbrace \,\vert \,T,
    \Delta, \bm{X}, A_k\big] }{{G}(T\! -  \, \vert \, \bm{X}, A_k)} \,
    \bigg\vert \, \bm{X}, A_k\right]
  \\
  &= \EE \left[\1 \lbrace {T} \le t_0 , \Delta= 1\rbrace \, \frac{
    {G}(T \! - \, \vert \, \bm{X}, A_k) }{{G}(T \! -  \, \vert \, \bm{X}, A_k)} \,
    \bigg\vert \, \bm{X}, A_k\right]
  \\
  &= \EE \big[\1 \lbrace {T} \le t_0 , \Delta= 1\rbrace \, \vert\, \bm{X},A_k\big]
\end{align*}
and
\begin{align*}
  \EE[ \1 \lbrace {T} \le t_0 , \Delta= 1\rbrace \, \vert \,  \bm{X}, A_k=a ]
  &= \EE[\1 \lbrace {T}^a \le t_0 , \Delta^a= 1\rbrace\, \vert \,  \bm{X}, A_k=a ] \\
&  =  \EE[\1 \lbrace {T}^a \le t_0 , \Delta^a= 1\rbrace\, \vert \,  \bm{X} ], 
\end{align*}
for \(a=0,1\), which yields \eqref{eq:crude:111}.

\section*{Appendix C}

\setcounter{section}{3}

To explain the general idea of GRFs, we use a generic (uncensored)
random variable \(Y\in \R\) and a corresponding generic parameter of
interest,
\begin{align*}
\theta(\bm{x})= \EE [Y \, \vert \, A_k=1, \bm{X}=\bm{x}] - \EE [Y \,
  \vert \, A_k=0, \bm{X}=\bm{x}],
\end{align*}
representing the treatment effect of \(A_k\) on \(Y\) conditional on
\(\bm{X}=\bm{x}\).  \citet[][Section 6]{athey2019generalized} consider
a conditional average partial effect estimation problem which they
formulate in terms of a structural model.  Below we demonstrate
\color{black} the equivalence of their setting with the counterfactual
formulation and show that the conditional average treatment effect
estimation problem considered here is a special case. In particular,
we show that the parameter \(\theta(\bm{x})\) can be identified in
terms of
\begin{align}
 \theta(\bm{x}) =
  \frac{
  \mathrm{cov} (A_k, Y \, \vert \, \bm{X}=\bm{x})
  }{
\mathrm{Var} (A_k \, \vert \, \bm{X}=\bm{x})
  }.
  \label{eq:estimator:1}
\end{align}
This means that \( \theta(\bm{x})\) can be estimated by providing
estimators for \(\mathrm{cov} (A_k, Y \, \vert \, \bm{X}=\bm{x})\) and
\(\mathrm{Var} (A_k \, \vert \, \bm{X}=\bm{x})\), respectively.  The
forest outputs weights that can be used to define such estimators as
follows. First, forest weights are obtained by averaging over the
neighborhoods \(L_b(\bm{x})\) defined by the trees, \(b=1,\ldots, B\),
\begin{align}
  \alpha_i(\bm{x}) = \frac{1}{B} \sum_{b=1}^B \alpha_{b,i}(\bm{x}), \qquad
  \text{where,} \quad
  \alpha_{b,i} (\bm{x}) = \frac{\1 \lbrace \bm{X}_i \in L_b(\bm{x})
  \rbrace}{
  \sum_{k=1}^n \1 \lbrace \bm{X}_k \in L_b(\bm{x})
  \rbrace}.
  \label{eq:forest:weights}
\end{align}
Then, the forest estimator \( \hat{\theta}_{\alpha} (\bm{x}) \) is given by,
\begin{align}
  \hat{\theta}_{\alpha} (\bm{x}) = \left( \sum_{i=1}^n \alpha_i(\bm{x}) \left(A_i -
  \bar{A}_{k, \alpha} \right)^2 \right)^{-1} \left( \sum_{i=1}^n \alpha_i(\bm{x})
  \left(A_i -
  \bar{A}_{k, \alpha} \right)  \left(Y_i -
  \bar{Y}_{\alpha} \right) \right).
  \label{eq:theta:estimator}
\end{align}
Here, \(\bar{A}_{k, \alpha} = \sum_{i=1}^n \alpha_i(\bm{x}) A_i\) and
\(\bar{Y}_{\alpha} = \sum_{i=1}^n \alpha_i(\bm{x}) Y_i\) are
estimators for the propensity score
\(\pi_k (\bm{x}) = \EE [A_k\, \vert\, \bm{X}=\bm{x}]\) and for
\(\EE [Y\, \vert\, \bm{X}=\bm{x}]\), respectively.  \citet[][Theorem 5
and Section 6]{athey2019generalized} provide conditions under which
\(\hat\theta_{\alpha}\) converges in distribution to a normal
distribution centered around the true \( {\theta} (\bm{x})\).  They
further propose an estimator \(\hat{\sigma}_n(\bm{x})\) for the
standard deviation of the asymptotic distribution.

A key part of the generalized random forest algorithm is the splitting
rule that targets specifically the estimation of the quantity of
interest \(\theta(\bm{x})\). Each split starts with a mother node
\(M\subset \mathcal{X}\), corresponding to a subset of
\(\mathcal{X}\), that is to be split into two daughter nodes
\(D_1 \cupdot D_2 = M\). For \(l=1,2\), let \(\hat{\theta}_{D_l}\) be
the daughter node local estimate of \(\theta (\bm{x})\) given by
\eqref{eq:theta:estimator} with
\(\alpha_i (\bm{x}) = \1\lbrace \bm{X}_i \in D_{l}\rbrace\) that
simply gives weight one to all samples falling in the respective
daughter node.  To derive their approximate criterion for picking good
splits, \citet{athey2019generalized} use a gradient-based
approximation of the mother node estimator \(\hat{\theta}_{M}\). In
the setting without censoring and competing risks, as we demonstrate
below\color{black}, it can be seen that the ``pseudo-outcomes'' used
in the ``labeling step'' of the splitting rule correspond to mother
node specific estimates of the efficient influence function for the
target parameter. Specifically, the split criterion is based on,
\begin{align}
  \rho_i  = W_M^{-1} (A_{k,i} - \bar{A}_M) \Big( Y_{i} - \bar{Y}_M - \big(A_{k,i} - \bar{A}_M\big)
  \hat{\theta}_M \Big),
  \label{eq:rho:i:0}
\end{align}
where,
\begin{align*}
  W_M = \frac{1}{\# \lbrace i \, : \, \bm{X}_i \in M\rbrace}
  \sum_{\lbrace i \, : \, X_i \in M\rbrace} (A_{k,i} - \bar{A}_M)^2,
\end{align*}
and \(\bar{A}_M\), \(\bar{Y}_{M}\) are mother node averages.
Each split of a mother node \(M\) into daughter nodes \(D_1, D_2\) is
carried out such as to maximize,
\begin{align*}
  \tilde{\mathcal{L}} (D_1, D_2) = \sum_{l=1}^2
  \frac{1}{ \# \lbrace i \, \,:\, \bm{X}_i \in D_l\rbrace}
  \Bigg( \sum_{i\in \lbrace i \, \,:\, X_i \in D_l\rbrace} \rho_i \Bigg)^2,
\end{align*}
with \(\rho_i\) as defined in \eqref{eq:rho:i:0}.

\subsection{Equivalence between counterfactual formulation and
  structural model formulation}

We demonstrate the equivalence of the setting of \citet[][Section
6]{athey2019generalized} with the counterfactual formulation and show
that the conditional average treatment effect estimation problem
considered in the main paper (Section 4) is a special case hereof.

Accordingly, we here consider observed data \(O=(\bm{X},A_k,Y)\),
\(\bm{X} \in \mathcal{X}\), \(A_k\in\lbrace 0,1\rbrace\) and
\(Y\in \R\) (uncensored). Further, let \(Y^1\) be the counterfactual
outcome that would have been observed under \(A_k=1\), and \(Y^0\) be
the counterfactual outcome that would have been observed under
\(A_k=0\). The consistency assumption states that
\begin{align}
  Y = A_kY^1 + (1-A_k) Y^0,
  \label{eq:simple:consistency}
\end{align}
and the exogeneity assumption (no unmeasured confounding) that
\((Y^1, Y^0) \independent A_k \, \vert \, \bm{X}\).  The conditional
treatment effect is defined as,
\begin{align*}
  \theta( \bm{x}) = \EE  [ Y^1 - Y^0 \, \vert\,  \bm{X}= \bm{x}] =
  \EE  [ Y \, \vert\, A_k=1, \bm{X}= \bm{x}]
  -  \EE  [ Y \, \vert\, A_k=0, \bm{X}= \bm{x}]. 
\end{align*}
The second equality follows under the exogeneity assumption together
with the consistency assumption.

Assume on the other hand that,
\begin{align}
  Y_i = a_i +  b_i A_k + \eps_i, 
  \label{eq:structural}
\end{align}
equivalent to \cite[][Section 6]{athey2019generalized} with our
\(a_i+\eps_i\) collapsed into just \(\eps_i\).

We show that \eqref{eq:structural} imposes no restriction when \(A_k\)
is binary.  Under consistency, we can express \(Y\) as,
\begin{align*}
Y &=  A_kY^1 + (1-A_k) Y^0 \\
  &=  A_kY^1 + (1-A_k) Y^0 + A_k \big( \EE[Y^1\, \vert\, \bm{X}] - \EE[Y^0\, \vert\, \bm{X}]  \big)
  -   A_k \EE[Y^1\, \vert\, \bm{X}] - ( 1-A_k) \EE[Y^0\, \vert\, \bm{X}]
     \\
  &\qquad\qquad\qquad\qquad\qquad\qquad\qquad\qquad\qquad\qquad\qquad\qquad
    \qquad\qquad\qquad\qquad\qquad   + \,\EE[Y^0\, \vert\, \bm{X}] \\
  &= \EE[Y^0\, \vert\, \bm{X}]
    +  A_k \big( \EE[Y^1\, \vert\, \bm{X}] - \EE[Y^0\, \vert\, \bm{X}]  \big)
    + A_k \big( Y^1 - \EE[Y^1\, \vert\, \bm{X}] +
    (1-A_k) \big( Y^0 - \EE[Y^0\, \vert\, \bm{X}]\big). 
\end{align*}
So if we let,
\begin{align*}
  a_i& := \EE[Y^0\, \vert\, \bm{X}_i], \\
  b_i &:= \EE[Y^1\, \vert\, \bm{X}_i]- \EE[Y^0\, \vert\, \bm{X}_i], \quad\text{and,}\\
  \eps_i &:= (1-A_{k,i}) \big( Y^0 - \EE[Y^0\,\vert\, \bm{X}_i]\big)
           +A_{k,i} \big( Y^1 - \EE[Y^1\,\vert\, \bm{X}_i]\big),
\end{align*}
we are back on the form in \eqref{eq:structural}.

Further note that,
\begin{align*}
  \EE[ \eps_i \, \vert \, A_k , \bm{X}]
  =  (1-A_{k,i}) \big( \EE[Y^0\,\vert\, \bm{X}_i] - \EE[Y^0\,\vert\, \bm{X}_i]\big)
  + A_{k,i} \big( \EE[Y^1\,\vert\, \bm{X}_i] - \EE[Y^1\,\vert\, \bm{X}_i]\big) = 0,
\end{align*}
so that,
\begin{align*}
  \EE[ Y_i  \,\vert\, A_k, \bm{X}] = \EE[Y^0\, \vert\, \bm{X}_i] +  \theta(\bm{x}) A_k .
\end{align*}

\subsection{Identification of the target parameter}

We demonstrate that,
\begin{align}
 \theta(\bm{x}) =  \EE [  Y \, \vert \, A_k=1, \bm{X}=\bm{x}] 
  - \EE [Y \, \vert \, A_k=0, \bm{X}=\bm{x}] =
  \frac{
  \mathrm{cov} (A_k, Y \, \vert \, \bm{X}=\bm{x})
  }{
\mathrm{Var} (A_k \, \vert \, \bm{X}=\bm{x}) 
  }. 
  \label{eq:relation:theta}
\end{align}
This follows since \(A_k\in\lbrace 0,1 \rbrace\), so that we have: 
\begin{align*}
  \mathrm{cov} (A_k, Y \, \vert \, \bm{X}=\bm{x})
  &= \EE [ A_k  Y \, \vert \, \bm{X}=\bm{x}] - \EE[A_k\, \vert \, \bm{X}=\bm{x}] \, \EE [Y\, \vert \, \bm{X}=\bm{x}] \\
  &= \EE [ A_k  Y \, \vert \, A_k=1, \bm{X}=\bm{x}] \,\pi_{k}(\bm{x}) \\
& \qquad\qquad\qquad  + \,  \EE [ A_k  Y \, \vert \, A_k=0, \bm{X}=\bm{x}]\, (1-\pi_{k}(\bm{x})) - \pi_{k}(\bm{x})\, \EE [Y\, \vert \, \bm{X}=\bm{x}] \\
  &= \EE [  Y \, \vert \, A_k=1, \bm{X}=\bm{x}] \,\pi_{k}(\bm{x}) - \pi_{k}(\bm{x})\, \big( \EE [Y \, \vert \, A_k=1,\bm{X}=\bm{x}]
    \,\pi_{k}(\bm{x})  \\
  &\qquad\qquad\qquad\qquad\qquad\qquad\qquad\qquad\,\,\,
    + \, \EE [Y \, \vert \, A_k=0, \bm{X}=\bm{x}]\, (1-\pi_{k}(\bm{x})) \big)\\
  &= \EE [  Y \, \vert \, A_k=1, \bm{X}=\bm{x}] \,\pi_{k}(\bm{x}) (1- \pi_{k}(\bm{x})) \\
  & \qquad\qquad\qquad\qquad\,\,\qquad\qquad\qquad    -\, \EE [Y \, \vert \, A_k=0, \bm{X}=\bm{x}]\, \pi_{k}(\bm{x}) (1-\pi_{k}(\bm{x})) \\
  &= \big(\EE [  Y \, \vert \, A_k=1, \bm{X}=\bm{x}] 
    - \EE [Y \, \vert \, A_k=0, \bm{X}=\bm{x}]\big)\, \pi_{k}(\bm{x}) (1-\pi_{k}(\bm{x})) ,\\  
  &= \big(\EE [  Y \, \vert \, A_k=1, \bm{X}=\bm{x}] 
    - \EE [Y \, \vert \, A_k=0, \bm{X}=\bm{x}]\big) \, \mathrm{Var} (A_k \, \vert \, \bm{X}=\bm{x}),\\
\end{align*}
which yields \eqref{eq:relation:theta}.  

\subsection{Influence function used for splitting}

The influence function used to split in the GRF algorithm for
estimation of treatment effects \citep[][Section
6]{athey2019generalized} is,
\begin{align}
  \rho_i  = W_M^{-1} (A_{k,i} - \bar{A}_M) \Big( Y_{i} - \bar{Y}_M - \big(A_{k,i} - \bar{A}_M\big)
  \hat{\theta}_M \Big),
  \label{eq:rho:i}
\end{align}
where,
\begin{align*}
  W_M = \frac{1}{\# \lbrace i \, : \, \bm{X}_i \in M\rbrace}
  \sum_{\lbrace i \, : \, X_i \in M\rbrace} (A_{k,i} - \bar{A}_M)^2, 
\end{align*}
and \(\bar{A}_M\), \(\bar{Y}_{M}\) are mother node averages.  Note
that \(\rho_i\) in \eqref{eq:rho:i} is a mother node specific
estimator for,
\begin{align}
  \begin{split}
    &\phi (Y, A_k) = \big( \mathrm{Var}(A_k \, \vert \, \bm{x}) \big)^{-1}
    \big(A_k - \pi_{k}(\bm{x})\big) \big( Y - \EE[ Y\, \vert \,
    \bm{X}=\bm{x}] - \big(A_k - \pi_{k}(\bm{x})\big) \theta(\bm{x})
    \big).
    \end{split}
    \label{eq:if:expression:2}
\end{align}
We here demonstrate that \(\phi (Y, A_k)\) in
\eqref{eq:if:expression:2} can also be written,
\begin{align}
  \phi (Y, A_k)
  =
  \left( \frac{A_{k}}{\pi_{k}(\bm{x})} - \frac{1-A_{k}}{1-\pi_{k}(\bm{x})}\right)
  \Big( Y - \EE [Y \, \vert \, A_{k}, \bm{X}=\bm{x}] \Big),
  \label{eq:if:expression:1}
\end{align}
which we recognize as the efficient influence function for estimation
of the parameter
\(\theta(\bm{x})= \EE [Y \, \vert \, A_k=1, \bm{X}=\bm{x}] - \EE [Y \,
\vert \, A_k=0, \bm{X}=\bm{x}]\)
\citep{scharfstein1999adjusting,rosenblum2011simple}.

First note that
\(\mathrm{Var}(A_k \, \vert \, \bm{x}) = (1- \pi_{k}
(\bm{x}))\pi_{k}(\bm{x})\) since \(A_k\) is binary. Next, by iterated
expectations, we have that,
\begin{align*}
  \EE [Y \, \vert \, \bm{X}=\bm{x} ] = \EE [Y \, \vert \, A_k=1,\bm{X}= \bm{x} ] \,
  \pi_{k} (\bm{x}) + \EE [Y \, \vert \, A_k=0,\bm{X}= \bm{x} ] \,(1- \pi_{k}
  (\bm{x})).
\end{align*}
Moreover, we can write
\((A_k - \pi_{k} ) = A_k\, (1- \pi_{k} ) + (1-A_k)\,\pi_{k} \).  Also
recall that
\(\theta_{\mathrm{net}}(\bm{x}) = \EE [Y \, \vert \, A_k=1,
\bm{X}=\bm{x}]-\EE [Y \, \vert \, A_k=0, \bm{X}=\bm{x}]\).

Now rewrite,
\begin{align*}
 \frac{ A_k}{\pi_{k} (\bm{x}) } -  \frac{ (1-A_k)}{1-\pi_{k}(\bm{x}) }
  &= \frac{A_k (1- \pi_{k} (\bm{x}) )}{\pi_{k} (\bm{x}) (1- \pi_{k} (\bm{x}) ) }
    - \frac{(1-A_k) \pi_{k} (\bm{x}) }{(1-\pi_{k} (\bm{x})) \pi_{k} (\bm{x}) } 
  \\
  &= \big( A_k - \pi_{k}(\bm{x}) \big) \, \big( \mathrm{Var}(A_k \, \vert \, \bm{x}) \big)^{-1},
\end{align*}
and,
\begin{align*}
  &\EE [Y \, \vert \, A_k=1, \bm{X}=\bm{x}] \\
  &\qquad = \pi_{k} (\bm{x}) \,  \EE [Y \, \vert \, A_k=1, \bm{X}=\bm{x}]
    + (1- \pi_{k}(\bm{x})) \, \EE [Y \, \vert \, A_k=1, \bm{X}=\bm{x}] \\
  &\qquad =   \EE [Y \, \vert \,  \bm{X}=\bm{x}] -
    \EE [Y \, \vert \, A_k=0,\bm{X} =\bm{x} ] (1- \pi_{k} (\bm{x}))
    + (1- \pi_{k} (\bm{x})) \, \EE [Y \, \vert \, A_k=1, \bm{X}=\bm{x}] \\
  &\qquad =   \EE [Y \, \vert \,  \bm{X}=\bm{x}] - (1- \pi_{k} (\bm{x})) \big(
    \EE [Y \, \vert \, A_k=0,\bm{X} =\bm{x} ] - \EE [Y \, \vert \, A_k=1, \bm{X}=\bm{x}]\big) \\
  &\qquad =   \EE [Y \, \vert \,  \bm{X}=\bm{x}] +(1 - \pi_{k} (\bm{x})) \big(
    \EE [Y \, \vert \, A_k=1,\bm{X} =\bm{x} ] - \EE [Y \, \vert \, A_k=0, \bm{X}=\bm{x}]\big) ,
\end{align*}  
and likewise,
\begin{align*}
  &\EE [Y \, \vert \, A_k=0, \bm{X}=\bm{x}] \\
  &\qquad = \pi_{k} (\bm{x}) \,  \EE [Y \, \vert \, A_k=0, \bm{X}=\bm{x}]
    + (1- \pi_{k}(\bm{x})) \, \EE [Y \, \vert \, A_k=0, \bm{X}=\bm{x}] \\
  &\qquad =   \EE [Y \, \vert \,  \bm{X}=\bm{x}] +
    \EE [Y \, \vert \, A_k=0,\bm{X} =\bm{x} ] \,\pi_{k} (\bm{x})
    -  \pi_{k}(\bm{x}) \, \EE [Y \, \vert \, A_k=1, \bm{X}=\bm{x}] \\
  &\qquad =   \EE [Y \, \vert \,  \bm{X}=\bm{x}] -\pi_{k} (\bm{x}) \big(
    \EE [Y \, \vert \, A_k=0,\bm{X} =\bm{x} ] - \EE [Y \, \vert \, A_k=1, \bm{X}=\bm{x}]\big) \\
  &\qquad =   \EE [Y \, \vert \,  \bm{X}=\bm{x}] + (0 - \pi_{k} (\bm{x})) \big(
    \EE [Y \, \vert \, A_k=1,\bm{X} =\bm{x} ] - \EE [Y \, \vert \, A_k=0, \bm{X}=\bm{x}]\big) .
\end{align*}  
Collecting the above, we rewrite \eqref{eq:if:expression:1} as, 
\begin{align*}
  \phi (Y, A_k) 
  &= \left(  \frac{A_k}{\pi_{k}(\bm{x})}
    -  \frac{1-A_k}{1-\pi_{k}(\bm{x})} \right) \Big( Y -   \EE [ Y \, \vert \, A_k, \bm{X}=\bm{x}]
    \Big) \\
  &= \big( A_k - \pi_{k} (\bm{x}) \big) \, \big( \mathrm{Var}(A_k \, \vert \, \bm{x}) \big)^{-1}
    \Big( Y - \EE[Y \, \vert \, \bm{X}=\bm{x}] - (A_k - \pi_{k}(\bm{x}) ) \, \theta(\bm{x}) \Big),
\end{align*}
which yields \eqref{eq:if:expression:2}.

\section*{Appendix D}

\setcounter{section}{4}

We here collect descriptive statistics for our data analysis.

\begin{figure}[!h]   
   \centering \includegraphics[width=0.7\textwidth,angle=0]
   {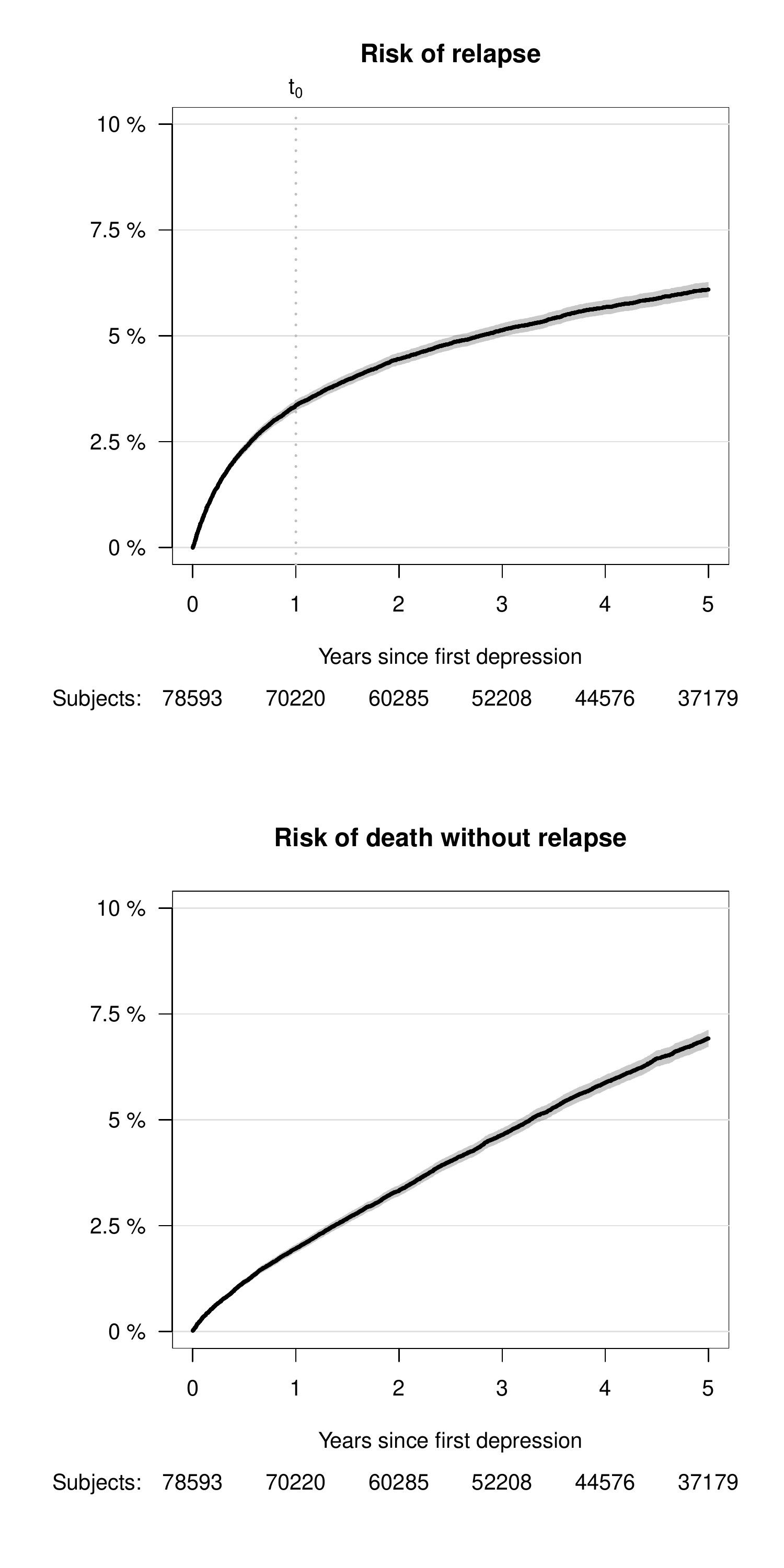}
   \caption{Aalen-Johansen estimators for the risk of readmission with
     depression (top) and risk of death without relapse (bottom),
     respectively. We are interested in readmissions within 1 year
     which is marked on the upper plot. The number of subjects at risk
     is shown below the plots.}
   \label{fig:plot:aalen:johansen} 
\end{figure}

Figure \ref{fig:plot:aalen:johansen} shows unadjusted Aalen-Johansen
estimators \citep{aalen1978empirical} for the risk of readmission with
depression and risk of death without relapse, respectively.

Table \ref{table:summary:statistics:1} shows the number of subjects in
each age group and in each comorbidity group. Table
\ref{table:summary:statistics:2} shows the number of subjects exposed
to the different drug groups in the exposure window. Table
\ref{table:no:of:events} shows the number of relapse with depression
within five years, along with number of subjects who die without
depression.

\begin{table}[ht]
\centering
\begin{tabularx}{\textwidth}{Lllll}
  & Male (n=28748) & Female (n=49952) & Total (n=78700) \\ 
  \midrule
Infections & 6429 (22.4) & 14183 (28.4) & 20612 (26.2) \\ 
  Neoplasms & 3775 (13.1) & 9419 (18.9) & 13194 (16.8) \\ 
  Diseases of blood & 674 (2.3) & 1403 (2.8) & 2077 (2.6) \\ 
  Diseases of the nervous system & 6560 (22.8) & 11836 (23.7) & 18396 (23.4) \\ 
  Diseases of the circulatory or respiratory system & 9446 (32.9) & 16408 (32.8) & 25854 (32.9) \\ 
  Nutritional and metabolic diseases & 6863 (23.9) & 11752 (23.5) & 18615 (23.7) \\ 
  Diseases of the skin and subcutaneous tissue & 2040 (7.1) & 3752 (7.5) & 5792 (7.4) \\ 
  Diseases of the musculoskeletal system & 7856 (27.3) & 15212 (30.5) & 23068 (29.3) \\ 
  Diseases of the genitourinary system and pregnancy, childbirth and the puerperium & 4002 (13.9) & 21003 (42.0) & 25005 (31.8) \\ 
  age in (0,18] & 1900 (6.6) & 4595 (9.2) & 6495 (8.3) \\ 
  age in (18,25] & 3437 (12.0) & 7466 (14.9) & 10903 (13.9) \\ 
  age in (25,30] & 2213 (7.7) & 4347 (8.7) & 6560 (8.3) \\ 
  age in (30,40] & 4668 (16.2) & 8457 (16.9) & 13125 (16.7) \\ 
  age in (40,50] & 5216 (18.1) & 7360 (14.7) & 12576 (16.0) \\ 
  age in (50,60] & 4349 (15.1) & 5321 (10.7) & 9670 (12.3) \\ 
  age in (60,70] & 2689 (9.4) & 3486 (7.0) & 6175 (7.8) \\ 
  age in (70,80] & 2308 (8.0) & 4054 (8.1) & 6362 (8.1) \\ 
  age $>$ 80 & 1968 (6.8) & 4866 (9.7) & 6834 (8.7) \\
   \bottomrule
\end{tabularx}
\caption{Comorbidities and demographics of the Danish population-based
  registry study. Shown are counts (\%).}
\label{table:summary:statistics:1}
\end{table}

\begin{table}[ht]
\centering
\begin{tabular}{lllll}
  & Male (n=28748) & Female (n=49952) & Total (n=78700) \\ 
  \midrule
  N06 & 18740 (65.2) & 33327 (66.7) & 52067 (66.2) \\ 
  N05 & 10049 (35.0) & 16784 (33.6) & 26833 (34.1) \\ 
  N02 & 3384 (11.8) & 7467 (14.9) & 10851 (13.8) \\ 
  A02 & 2509 (8.7) & 4486 (9.0) & 6995 (8.9) \\ 
  J01 & 2118 (7.4) & 5739 (11.5) & 7857 (10.0) \\ 
  B01 & 2687 (9.3) & 3484 (7.0) & 6171 (7.8) \\ 
  N03 & 1713 (6.0) & 2965 (5.9) & 4678 (5.9) \\ 
  C03 & 1610 (5.6) & 3450 (6.9) & 5060 (6.4) \\ 
  G03 & 42 (0.1) & 7352 (14.7) & 7394 (9.4) \\ 
  R03 & 1254 (4.4) & 2381 (4.8) & 3635 (4.6) \\ 
  C09 & 2219 (7.7) & 3079 (6.2) & 5298 (6.7) \\ 
  M01 & 1503 (5.2) & 3082 (6.2) & 4585 (5.8) \\ 
  C10 & 1822 (6.3) & 2356 (4.7) & 4178 (5.3) \\ 
  A10 & 1236 (4.3) & 1339 (2.7) & 2575 (3.3) \\ 
  C07 & 1401 (4.9) & 2075 (4.2) & 3476 (4.4) \\ 
  S01 & 866 (3.0) & 2149 (4.3) & 3015 (3.8) \\ 
  C08 & 1170 (4.1) & 1897 (3.8) & 3067 (3.9) \\ 
  A12 & 815 (2.8) & 1907 (3.8) & 2722 (3.5) \\ 
  A06 & 756 (2.6) & 1503 (3.0) & 2259 (2.9) \\ 
  C01 & 695 (2.4) & 1082 (2.2) & 1777 (2.3) \\ 
  G04 & 1107 (3.9) & 328 (0.7) & 1435 (1.8) \\ 
  H03 & 235 (0.8) & 1390 (2.8) & 1625 (2.1) \\ 
  D07 & 633 (2.2) & 1234 (2.5) & 1867 (2.4) \\ 
  N07 & 897 (3.1) & 666 (1.3) & 1563 (2.0) \\ 
  B03 & 485 (1.7) & 1076 (2.2) & 1561 (2.0) \\ 
  R05 & 370 (1.3) & 951 (1.9) & 1321 (1.7) \\ 
  R06 & 442 (1.5) & 1143 (2.3) & 1585 (2.0) \\ 
  A03 & 334 (1.2) & 1010 (2.0) & 1344 (1.7) \\ 
  M05 & 158 (0.5) & 1011 (2.0) & 1169 (1.5) \\ 
  H02 & 374 (1.3) & 755 (1.5) & 1129 (1.4) \\ 
  N04 & 261 (0.9) & 431 (0.9) & 692 (0.9) \\ 
  D01 & 458 (1.6) & 708 (1.4) & 1166 (1.5) \\ 
  R01 & 357 (1.2) & 672 (1.3) & 1029 (1.3) \\ 
  \bottomrule
\end{tabular}
\caption{Number (\%) of subjects purchasing treatments in the Danish
  population-based registry study.}
\label{table:summary:statistics:2}
\end{table}

\begin{table}[ht]
\centering
\begin{tabular}{llll}
  $\Delta$ & event type & number of subjects & percent of total \\ 
  \midrule
  0 & censoring & 67794 & 86.14 \% \\ 
  1 & depression relapse & 4613 & 5.861 \% \\
  2 & competing event & 6293 & 7.996 \% \\ 
  \bottomrule
\end{tabular}
\caption{Number of relapse events, censoring and competing events
  after five years.  }
\label{table:no:of:events}
\end{table}

To illustrate the effects of covariates for the estimation of our
target parameters, we compare our forest estimates of
\(\bar{\theta}_{\mathrm{crude}}\) to the naive Aalen-Johansen estimates of crude
probabilities stratified on each treatment variable (that leaves out
all covariate information), i.e., the nonparametric and unadjusted
estimator of,
\begin{align*}
  P( T \le t_0, \Delta=1 \, \vert \, A_k =1 ) - P( T\le t_0 , \Delta=1 \, \vert \, A_k =0 ).
\end{align*}

\begin{figure}[!h]   
  \centering
  \makebox[\textwidth][c]{\includegraphics[width=1.2\textwidth,angle=0]
    {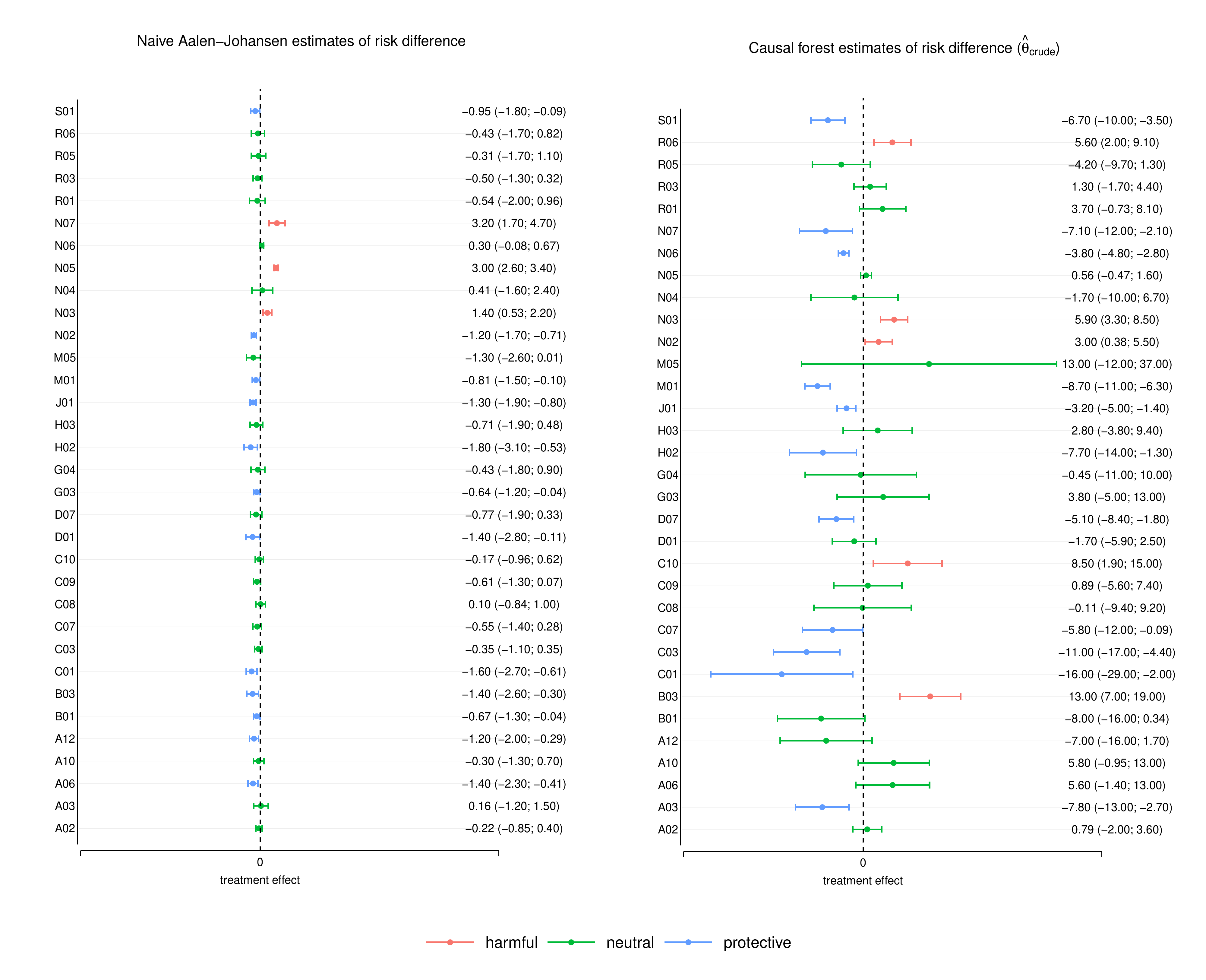}}
\caption{\textit{Left:} Naive Aalen-Johansen estimates of the risk
  difference, i.e., the difference in crude probabilities stratified
  on the respective treatment. \textit{Right:} Causal forest estimates
  of the effect on crude probabilities,
  \(\bar{\theta}_{\mathrm{crude}}\) (using unadjusted weights to
  construct weighted outcomes). For each ATC group (marked on the
  \(y\)-axis) the plot shows the estimates and the estimated
  confidence intervals (numbers written on the right). The colors
  indicate the direction of the effect. }
  \label{fig:plot1} 
\end{figure}

The naive Aalen-Johansen estimates along with confidence intervals and
the corresponding causal forest estimates of the treatment effect on
the crude probabilities, \(\bar{\theta}_{\mathrm{crude}}\), are shown
in Figure \ref{fig:plot1}. We see that the treatment effect estimates
for some drug groups differ quite a lot for the two methods.
Considering these differences, we deduce that there is a covariate
effect to be taken into account.

\label{lastpage}


\end{document}